\documentclass[11pt]{article}

\usepackage{cite}
\usepackage{amsmath,amssymb}
\usepackage{authblk}
\usepackage{bm}
\usepackage{lineno}
\usepackage{subcaption}
\usepackage{mwe}
\usepackage{setspace}
\usepackage{algorithm}
\usepackage{algpseudocode}
\usepackage{caption}

\usepackage{array}
\usepackage{subdepth}
\usepackage{url}
\usepackage{fullpage} 
\usepackage[colorlinks=true,citecolor=blue,linkcolor=blue,urlcolor=blue]{hyperref}

\newcommand{\bhline}{\noalign{\hrule height 0.9pt}}

\title{Events Meet Phase-Shifting Digital Holography:\\ Practical Acquisition, Theory, and Algorithms}

\author[1]{Ittetsu Uchiyama}
\author[1\footnote{ctsutake@nagoya-u.jp}]{Chihiro Tsutake}
\author[1]{Keita Takahashi}
\author[1]{Toshiaki Fujii}

\affil[1]{
Department of Information and Communication Engineering, 
Nagoya University, 
Furo-cho, 
Chikusa-ku, 
Nagoya, 
464-8603, 
Japan}

\date{\empty}

\begin{document}

\maketitle

\vspace{-10mm}

\begin{abstract}
We introduce a novel phase-shifting digital holography~(PSDH) method leveraging a hybrid event-based vision sensor~(EVS).
The key idea of our method is the phase shift during a single exposure.
The hybrid EVS records a hologram blurred by the phase shift, 
together with the events corresponding to blur variations.
We present analytical and optimization-based methods that theoretically support the reconstruction of full-complex wavefronts from the blurred hologram and events.
The experimental results demonstrate that our method achieves a reconstruction quality comparable to that of a conventional PSDH method while enhancing the acquisition efficiency.
\end{abstract}

\section{Introduction}
\label{s1}
Wavefront reconstruction is a fundamental problem in optics that has attracted significant attention across diverse applications~\cite{Gigan:22}.
In light of the physical limitations of sensors, 
various optical and computational methods~\cite{Goodman:67,Fienup:82,Streibl:84,Platt:01} have been developed to reconstruct complex wavefronts from constrained data.
Among these, we focus on digital holography~\cite{Goodman:67},
which captures intensity frames as holograms via an interferometric measurement.
Due to the large pixel size of sensors,
interference fringe spacing should be as large as possible to avoid aliasing noise.
While in-line holography has the capability of maximizing the fringe spacing, 
it suffers from zero-order and twin images~\cite{Stoykova:14} in the reconstructed wavefronts.

\begin{figure}[!t]
\centering
\includegraphics[scale=1.1]{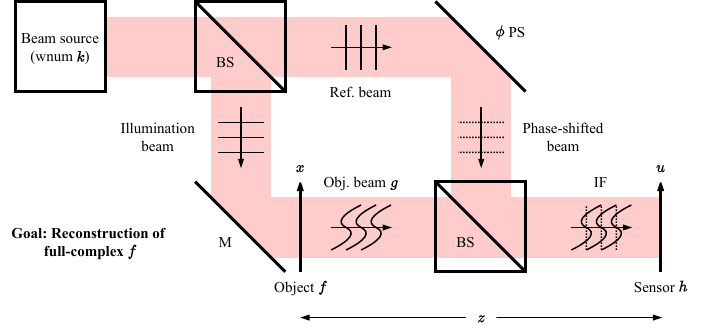}
\caption{Phase-shifting digital holography~(PSDH). Red region: optical path. BS: beam splitter. M: mirror. PS: phase shifter. IF: interference fringe. $k$: wavenumber. $\phi$: phase-shifting value. Conventional method: ordinary image sensor. Our method: hybrid event-based vision sensor~(EVS).}
\label{fig:psdh}
\end{figure}

\begin{figure}[t]
\centering
\begin{subfigure}{0.999\textwidth}
\centering
\includegraphics[scale=1.1]{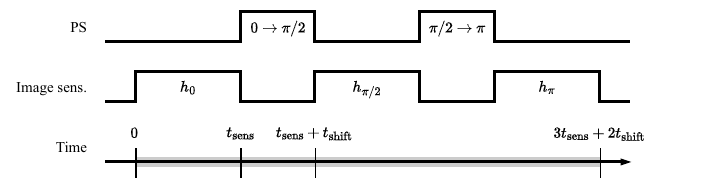}
\caption{Conventional frame-only PSDH~(F-PSDH) with image sensor. Three exposures with discrete phase shifts of  $0 \rightarrow \pi/2$ and $\pi/2 \rightarrow \pi$.}
\label{fig:tc_a}
\vspace{5mm}
\end{subfigure}
\begin{subfigure}{0.999\textwidth}
\centering
\includegraphics[scale=1.1]{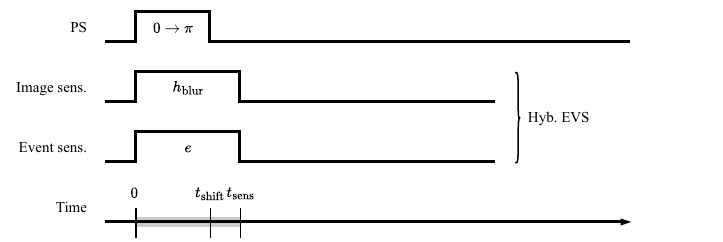}
\caption{Proposed frame-and-event PSDH~(FE-PSDH) with hybrid EVS. Single exposure with continuous phase shift of $0\rightarrow\pi$. Key idea: phase shift during single exposure.}
\label{fig:tc_b}
\end{subfigure}
\caption{Timing charts of acquisition. $t_\mathrm{sens}$: exposure time of sensor. $t_\mathrm{shift}$: transient time of phase shifter. Gray bar: total acquisition time. Phase-shifting value is initialized to $0$ before the first exposure. Throughout this paper, we assume $t_\mathrm{shift} \leq t_\mathrm{sens}$ and both methods share $t_\mathrm{sens}$ and $t_\mathrm{shift}$. Acquisition time of conventional method: $3t_\mathrm{sens}+2t_\mathrm{shift}$. Our method: $t_\mathrm{sens}$---single exposure time alone.}
\label{fig:tc}
\end{figure}

Phase-shifting digital holography~(PSDH)~\cite{Yamaguchi:97} is one of the methods for eliminating the unwanted images.
Figure~\ref{fig:psdh} shows a schematic illustration of PSDH.
A phase shifter modulates the global phase of the reference beam and an image sensor records a hologram, 
which is the intensity of interference fringes formed between the reference and object beams.
Figure~\ref{fig:tc_a} shows a timing chart of acquisition.
In PSDH, recording a hologram for each shifting value is essential, 
but it is highly time-consuming.
Parallel digital holography~\cite{Awatsuji:04,Awatsuji:06,Tahara:10} can resolve this temporal issue by spatially multiplexing holograms in a single frame.
However, the parallel methods sacrifice the spatial resolution of holograms,
thereby degrading the quality of the reconstructed wavefronts.

To address the temporal issue without sacrificing the spatial resolution,
we introduce a novel PSDH method leveraging a hybrid event-based vision sensor~(EVS)~\cite{Lichtsteiner:08},
which can record absolute intensity frames and relative intensity changes, i.e., events, in a logarithmic scale.
The key idea of our method is the phase shift of the reference beam during a single exposure, as illustrated in Fig.~\ref{fig:tc_b}.
The hybrid EVS records a hologram blurred by the phase shift, 
together with the events corresponding to blur variations.
We present a closed-form representation of the object wavefront in a noiseless setup,
which can be computed from the blurred hologram and events.
We also propose an optimization-based method to stabilize the wavefront reconstruction.
The experimental results demonstrate that our method achieves a reconstruction quality comparable to that of a conventional PSDH method while enhancing the acquisition efficiency.
\vspace{2mm}

\noindent
\textbf{Remark~1.} Our method requires a hybrid EVS, 
whose cost and availability are currently more limited than those of conventional image sensors. 
In addition, the relatively large pixel size of current EVS devices can limit the accurate recording of interference fringes, which poses a challenge for holographic recording.
Therefore, this work is positioned as a proof-of-concept study that demonstrates the feasibility of reducing the acquisition time in PSDH using event-based sensing.

\section{Related work}
\label{s2}
Event-based vision sensors have gained significant attention across a wide range of research fields, 
including computer vision.
Fu et al.~\cite{Fu:23} and Yu et al.~\cite{Yu:24} proposed three-dimensional reconstruction methods,
where they combined an EVS with illumination modulation for efficient data acquisition.
Habuchi et al.~\cite{Habuchi:24} proposed a light-field imaging method,
integrating coded aperture modulation with an EVS.
Inspired by these ideas, our PSDH leverages an EVS to capture continuously modulated (phase shifted) optical information within a single acquisition.
However, applying the optical modulation ideas of \cite{Fu:23,Yu:24,Habuchi:24} to PSDH is non-trivial due to fundamental differences in measurement models and reconstruction objectives.
This work is positioned as an investigation of the feasibility of single-exposure PSDH using an EVS.

Recently, event-based sensing has been actively explored in digital holographic recording, 
particularly for addressing motion-induced degradation~\cite{Uchiyama:24,Wang:25,Ge:25}.
However, existing event-driven holography methods primarily focus on motion-blur compensation and rely on non-phase-shifting hologram recording.
As a result, the twin-image problem remains fundamentally unresolved in these methods.
Moreover, phase information is not explicitly recovered through event measurements.
In contrast, our work explores a different direction by integrating event-based sensing with PSDH,
enabling twin-image-free reconstruction under practical acquisition.
To the best of our knowledge, 
we are the first to provide a systematic perspective on PSDH integrated with event-based sensing.

\section{Preliminaries}
\label{s3}
\textbf{Terminologies}.
The conventional PSDH~\cite{Yamaguchi:97} that only uses frames~(non-blurred holograms) is referred to as \textit{F-PSDH}.
The proposed method that incorporates a single frame~(blurred hologram) and events is referred to as \textit{FE-PSDH}.
\vspace{2mm}

\noindent
\textbf{Optics}.
As shown in Fig.~\ref{fig:psdh},
without loss of generality, we consider one-dimensional object and sensor planes.
The complex wavefront on the object plane is denoted by $f$.
The intensity distribution on the sensor plane, i.e., the hologram, is denoted by $h$.
We assume that $f$ and $h$ are parallel and located at a distance $z$ from each other.
The coordinates on the object and sensor planes are denoted by $x$ and $u$, respectively.
For notational simplicity, 
we sometimes omit the coordinates when they can be inferred from the context.
The complex wavefront of the object beam is denoted by $g$.
As shown in Fig.~\ref{fig:tc}, we assume that F-PSDH and FE-PSDH share the same exposure time for capturing a single frame,
which is denoted by $t_\mathrm{sens}$.
The operating time of a phase shifter is denoted by $t_\mathrm{shift}$.
We assume $t_\mathrm{shift} \leq t_\mathrm{sens}$.
The wavenumber of a beam is denoted by $k$.
The phase-shifting value is denoted by $\phi$.
We set $\phi=0$ before acquisition.
\vspace{2mm}

\noindent
\textbf{Mathematics.}
The imaginary unit is denoted by $j$.
The amplitude of a real or complex number is denoted by $|\cdot|$.
The complex conjugate is denoted by $\bar{\cdot}$.
The Euclidean norm of a function is denoted by $\|\cdot\|_2$.

\section{Principle of F-PSDH}
\label{s4}
In this section, 
we mention a general method for reconstructing the object wavefront $f$ from the intensity frames obtained by an ordinary image sensor.
The complex wavefront of the object beam at the sensor plane, i.e., $g$, 
is represented as the wave propagated from $f$ based on the Fresnel diffraction integral~\cite{Goodman05}.
Let $r$ be the complex wavefront of the reference beam at the sensor plane.
As shown in Fig.~\ref{fig:psdh}, 
a phase shifter modulates the global phase of $r$, resulting in $r\exp(j\phi)$.
A digital hologram, i.e.,
the intensity of interference fringes formed between $g$ and $r\exp(j\phi)$, can be formulated as
%
\begin{align}
h_\phi=|g+r\exp({j\phi})|^2=|g|^2+|r|^2+\bar{g}r\exp({j\phi})+g\bar{r}\exp({-j\phi}).
\label{eq:3_holo}
\end{align}
The terms $|g|^2+|r|^2$ and $\bar{g}r\exp(j\phi)$ are referred to as the zero-order and twin images, respectively.
To eliminate these unwanted images, the three-step method~\cite{Yamaguchi:97} is used here,
which records $h_0$, $h_{\pi/2}$, $h_\pi$ with three exposures.
By combining the three holograms, the complex wavefront $g$ can be perfectly reconstructed as
\begin{equation}
g=\frac{1-j}{4\bar{r}}\bigg(h_0-h_{\pi/2} + j(h_{\pi/2}-h_\pi)\bigg).
\label{eq:3_psdh}
\end{equation}
The complex object wavefront $f$ can be obtained from $g$ and the reverse Fresnel diffraction integral.
In conventional F-PSDH, alternating the phase shift and frame recording is essential, 
which increases acquisition time.
Specifically, if the operations of a phase shifter and an image sensor are synchronized,
the three-step method requires $3t_\mathrm{sens}+2t_\mathrm{shift}$, as shown in Fig.~\ref{fig:tc_a}.

\section{Proposed method: FE-PSDH}
\label{s5}
\subsection{Data acquisition}
\label{s5ss1}
To efficiently acquire data for reconstructing $f$ within a practical time,
we replace the conventional image sensor with the hybrid EVS~\cite{Lichtsteiner:08}.
It is an extension of the image sensor,
in which each pixel has a standard photon-detecting element with a bio-inspired event-sensing circuit.
This design allows the EVS to simultaneously capture absolute light intensity and detect changes in log light intensity.
As illustrated in Fig.~\ref{fig:tc_b}, we parallelize the phase shift with the frame and event recording,
where the phase of the reference beam is modulated from $0$ to $\pi$ during a single exposure.
With this design, the EVS records a hologram blurred by the phase shift,
which is denoted by $h_\mathrm{blur}$.
At the same time, the events triggered by blur variations are obtained asynchronously,
which are denoted by $e$.
Notably, as shown in Fig.~\ref{fig:tc_b}, 
the acquisition time can be reduced to $t_\mathrm{sens}$, the single exposure time.

\subsection{Analytical reconstruction in noiseless setup}
\label{s5ss2}
In this sub-section,
we show that $f$ can be recovered from the ideal noiseless $h_\mathrm{blur}$ and $e$.
The analytical reconstruction presented here is based on established Fresnel diffraction and F-PSDH theory. 
Our contribution is not the development of a new mathematical framework, 
but the first analytical justification of single-exposure PSDH assisted by event data.
Before deriving a reconstruction method,
we define mathematical models of the frame and event recording.
\vspace{2mm}

\noindent
\textbf{Frame recording model.}
Let $t=0$ be the beginning time of exposure and the synchronized phase shift,
$T_\mathrm{frame} = [0,t_\mathrm{sens}]$ be a continuous time duration for frame exposure,
and $\phi(t)$ be the phase-shifting value during its transition at time $t \in T_\mathrm{frame}$.
The noiseless frame $h_\mathrm{blur}$ can be modeled as the integration of non-blurred holograms in \eqref{eq:3_holo} over $t$ as follows:
\begin{equation}
h_\mathrm{blur}=\int_{T_\mathrm{frame}} h_{\phi(t)}dt 
= \lim_{M\rightarrow\infty}\frac{1}{M}\sum_{m=1}^{M}h_{\phi(t_\mathrm{sens}m/M)},
\label{eq:4_2_blur1}
\end{equation}
where $h_{\phi(t)}$ represents the hologram at the phase-shifting value $\phi = \phi(t)$.
Equation~\eqref{eq:4_2_blur1} assumes a time-invariant object; therefore, it is not strictly valid when object motion occurs during the exposure.
However, as demonstrated in Section~\ref{s6}, 
our method can provide stable and visually acceptable reconstructions for small object motions.
\vspace{2mm}

\noindent
\textbf{Event recording model.}
During the exposure, 
the EVS records the discrete time, spatial coordinate, and polarity of intensity changes.
Let $T_\mathrm{event} \subseteq T_\mathrm{frame}$ be a set of recorded discrete times during the exposure.
Events can be viewed as a set of signed delta functions, as
\begin{equation}
e(t,u)=\pm \delta\Big(u-u(t)\Big),
\label{eq:4_2_delta}
\end{equation}
where $u(t)$ is the spatial coordinate of the event recorded at the time $t \in T_\mathrm{event}$.
An event is triggered when the intensity change exceeds a contrast threshold~\cite{Lichtsteiner:08},
denoted by $C$, i.e.,
\begin{equation}
\Big|\log \Big(h_{\phi(t)}(u)\Big)-\log \Big(h_{\phi(t_\mathrm{prev})}(u)\Big)\Big| > C,
\label{eq:4_2_thresh}
\end{equation}
where $t_\mathrm{prev}$ is the time when the previous event was recorded.
We accumulate events from $t=0$ to a prescribed time $t$ for each $u$ as follows:
\begin{equation}
E_{0\rightarrow\phi(t)}(u)=\sum_{\tau} e(\tau,u) \quad \mathrm{s.t.} \quad 
\tau \in T_\mathrm{event} \quad \mathrm{and} \quad \tau \leq t.
\label{eq:4_2_acc}
\end{equation}
%
%
Equations~\eqref{eq:4_2_thresh} and \eqref{eq:4_2_acc} lead to the following approximate relationship;
\begin{equation}
E_{0\rightarrow\phi(t)} = \frac{1}{C}\Big(\log h_{\phi(t)}-\log h_0\Big),
\label{eq:4_2_evt_rel}
\end{equation}
where $h_0=h_{\phi(0)}$ and $\phi(0)=0$.
\vspace{2mm}

We define $\mathcal{E}$ as the integration of $E_{0\rightarrow\phi(t)}$ over the continuous time $t \in T_\mathrm{frame}$ as follows:
\begin{equation}
\mathcal{E} = 
\int_{T_\mathrm{frame}} \exp\Big(C E_{0\rightarrow\phi(t)}\Big)dt=
\lim_{N\rightarrow\infty}\frac{1}{N}\sum_{n=1}^{N}\exp\Big(C E_{0\rightarrow\phi(t_\mathrm{sens}n/N)}\Big).
\label{eq:4_2_vare}
\end{equation}
By transforming \eqref{eq:4_2_evt_rel}, we can obtain
\begin{equation}
h_{\phi(t)}=h_0\exp\Big(C E_{0\rightarrow{\phi(t)}}\Big).
\label{eq:4_2_hphi}
\end{equation}
Combining \eqref{eq:4_2_blur1}, \eqref{eq:4_2_vare}, and \eqref{eq:4_2_hphi}, we have
\begin{equation}
h_\mathrm{blur}=
\int_{T_\mathrm{frame}}h_0\exp\Big(C E_{0\rightarrow\phi(t)}\Big)dt = \mathcal{E}h_0,
\quad \therefore h_0(u) = \frac{h_\mathrm{blur}(u)}{\mathcal{E}(u)}.
\label{eq:4_2_blur2}
\end{equation}
From \eqref{eq:3_psdh}, \eqref{eq:4_2_evt_rel}, and \eqref{eq:4_2_blur2},
the object wavefront at the hologram plane, i.e., $g$, can be reconstructed as follows:
\begin{align}
g&=\frac{1-j}{4\bar{r}}\cdot h_0
\bigg\{1-\frac{h_{\pi/2}}{h_0} + 
j\bigg(\frac{h_{\pi/2}}{h_0}-\frac{h_\pi}{h_0}\bigg)\bigg\}\\
&=\frac{1-j}{4\bar{r}}\cdot \frac{h_\mathrm{blur}}{\mathcal{E}}
\bigg\{1-\exp(C E_{0\rightarrow\pi/2}) + 
j\bigg(\exp(CE_{0\rightarrow\pi/2})-
\exp(C E_{0\rightarrow\pi})\bigg)\bigg\}.
\label{eq:4_2_g}
\end{align}
Note that the right-hand side can be computed from the noiseless $h_\mathrm{blur}$ and $e$.
We can obtain the full-complex wavefront at the object plane, i.e., $f$, 
from $g$ and the reverse Fresnel diffraction integral, as was done in F-PSDH.

\subsection{Stable reconstruction via joint estimation}
\label{s5ss3}
To compute $E_{0\rightarrow\pi/2}$ in \eqref{eq:4_2_g},
it is essential to determine a time when $\phi(t) = \pi/2$, hereafter denoted by $t_{\pi/2}$.
In the noiseless setup,
this task is straightforward because $t_{\pi/2}$ can be computed from the response characteristic of the phase shifter.
In the practical noisy case, 
$h_\mathrm{blur}$ and $e$ are affected by various sources of noise,
such as pixel-wise fluctuations of the contrast threshold~\cite{Hu:21}.
For such noisy data,
we find that the exact $t_{\pi/2}$ value does not always present the best reconstruction,
as will be demonstrated in Section~\ref{s6ss1}.
Taking this background into account, 
we introduce an optimization-based reconstruction method that stably presents the accurate wavefront reconstruction in noisy conditions.
In the following,
the noisy versions of $h_\mathrm{blur}$ and $e$ are denoted by $\tilde{h}_\mathrm{blur}$ and $\tilde{e}$, respectively.

We first formulate an optimization problem to jointly recover $h_0$ and the time $t_{\pi/2}$.
Let us define the object wavefront at the sensor plane, parameterized by variables $h_{0,\mathrm{opt}}$~(real function) and $t_{\pi/2,\mathrm{opt}}$~(real number), as follows:
\begin{align}
&\tilde{g}\Big({h_{0,\mathrm{opt}},t_{\pi/2,\mathrm{opt}}}\Big)=\nonumber\\
&\frac{1-j}{4\bar{r}}\cdot h_{0,\mathrm{opt}}
\bigg\{1-\exp\Big(C \tilde{E}_{0\rightarrow\phi(t_{\pi/2,\mathrm{opt}})}\Big) + 
j\bigg(\exp\Big(C\tilde{E}_{0\rightarrow\phi(t_{\pi/2,\mathrm{opt}})}\Big)-
\exp\Big(C \tilde{E}_{0\rightarrow\pi}\Big)\bigg)\bigg\},
\end{align}
where $\tilde{E}_{0\rightarrow\phi(t_{\pi/2,\mathrm{opt}})}$ and $\tilde{E}_{0\rightarrow\pi}$ are computed by replacing $e$ in \eqref{eq:4_2_acc} with $\tilde{e}$.
We define the following two loss functions based on \eqref{eq:3_holo} and \eqref{eq:4_2_blur2}:
\begin{align}
&L_1\Big(h_{0,\mathrm{opt}},t_{\pi/2,\mathrm{opt}}\Big)=
\Big\| h_{0,\mathrm{opt}}-\Big|\tilde{g}\Big({h_{0,\mathrm{opt}},t_{\pi/2,\mathrm{opt}}}\Big)+r\Big|^2\Big\|_2^2,\\
&L_2\Big({h_{0,\mathrm{opt}}}\Big)=\|\tilde{\mathcal{E}}h_{0,\mathrm{opt}}-\tilde{h}_\mathrm{blur}\|_2^2,
\label{eq:loss}
\end{align}
where $\tilde{\mathcal{E}}$ is computed by replacing $E_{0\rightarrow\phi(t)}$ in \eqref{eq:4_2_vare} with $\tilde{E}_{0\rightarrow\phi(t)}$.
We formulate the optimization problem of the form
\begin{equation}
\min_{h_{0,\mathrm{opt}},t_{\pi/2,\mathrm{opt}}} \quad 
L\Big({h_{0,\mathrm{opt}},t_{\pi/2,\mathrm{opt}}}\Big)=
L_1\Big({h_{0,\mathrm{opt}},t_{\pi/2,\mathrm{opt}}}\Big)+\lambda L_2\Big({h_{0,\mathrm{opt}}}\Big),
\label{eq:4_3_argmin}
\end{equation}
where $\lambda$ is a positive hyperparameter.

We then solve the above problem using gradient descent, as
\begin{align}
&h_{0,\mathrm{opt}}[i+1]=
h_{0,\mathrm{opt}}[i]-\alpha_h\nabla
L\Big({h_{0,\mathrm{opt}}[i],t_{\pi/2,\mathrm{opt}}[i]}\Big),\\ 
&t_\mathrm{\pi/2,\mathrm{opt}}[i+1]=
t_\mathrm{\pi/2,\mathrm{opt}}[i]-\alpha_t\frac{\partial}{\partial t_\mathrm{\pi/2,\mathrm{opt}}}L\Big({h_{0,\mathrm{opt}}[i],t_{\pi/2,\mathrm{opt}}[i]}\Big),
\label{eq:4_3_gd}
\end{align}
where $\nabla$ is the gradient operator with respect to $h_{0,\mathrm{opt}}$, $i$ is the number of the current iteration, $h_{0,\mathrm{opt}}[i]$ and $t_{\pi/2,\mathrm{opt}}[i]$ are the variables at $i$,
and $\alpha_h$ and $\alpha_t$ are learning rates.
The gradient descent updates are repeated until the index $i$ reaches the predefined maximum number of iterations $I$.
Let $\tilde{f}$ denote the complex object wavefront reconstructed from noisy data.
In the optimization-based method, 
we obtain $\tilde{f}$ from $g = \tilde{g}\big({h_{0,\mathrm{opt}}[I],t_{\pi/2,\mathrm{opt}}[I]}\big)$ and the  reverse Fresnel diffraction integral.

\subsection{Implementation details}
\label{s5ss4}
\begin{figure}[!t]
\centering
\includegraphics[scale=1.1]{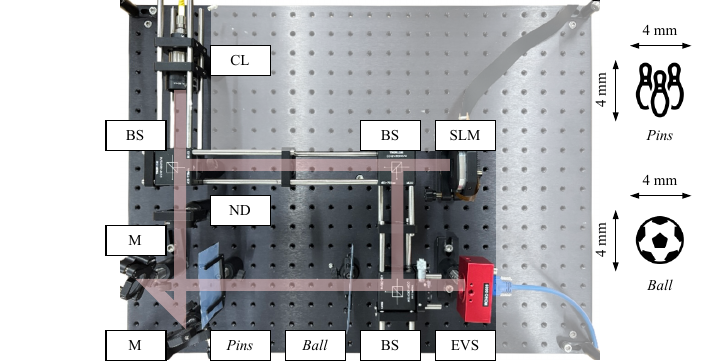}
\caption{Optical implementation of acquisition system in Fig.~\ref{fig:psdh}. CL: collimator lens. SLM: spatial light modulator. ND: neutral-density filter. \textit{Ball} and \textit{Pins} are printed on OHP sheets. EVS--\textit{Ball}: $120$~mm. EVS--\textit{Pins}: $270$~mm.}
\label{fig:impl}
\end{figure}

\textbf{Optics.}
Figure~\ref{fig:impl} illustrates an optical implementation of the acquisition system in Fig.~\ref{fig:psdh}.
The beam source was Thorlabs KLS635 of the wavelength $635$~nm; 
thus, the wavenumber was $k=2\pi/635$~rad/nm.
The collimator lens converts the incident beam into the front-parallel beam;
thus, the amplitude of $r$ was constant.
Santec SLM-210, a reflection-type spatial light modulator~(SLM), serves as a phase shifter,
which can modulate the phase of the incident reference beam pixel by pixel.
We placed an iniVation DAVIS346, a hybrid EVS, at the sensor plane,
which has the capability of recording monochrome frames and events.
The specifications of the phase shifter and hybrid EVS are summarized in Table~\ref{tab:spec}.
To reconstruct the full-complex wavefronts of three-dimensional objects,
the \textit{pins} and \textit{ball} patterns were printed on OHP sheets and placed at distances of $z=270$~mm and $120$~mm from the EVS, respectively.
The sizes of the printed objects were about $4 \times 4$~mm.
\vspace{2mm}

\begin{table*}[!t]
\centering
\renewcommand{\arraystretch}{1.4}
\caption{Specifications of phase shifter~(SLM-210) and hybrid EVS~(DAVIS346).}
\label{tab:spec}
\footnotesize
\begin{tabular}{cwc{18mm}wc{18mm}wc{18mm}}\bhline
Model              & \#pixels        & Pixel size~[\textmu m] & Time~[s]\\\hline
Santec SLM-210     & $1920 \times 1200$ & $ 8.0\times 8.0$     & $t_\mathrm{shift}=1/60$\\
iniVation DAVIS346 & $346 \times 260$   & $18.5 \times 18.5$   & $t_\mathrm{sens}=1/25$\\\bhline
\end{tabular}
\end{table*}

\noindent
\textbf{Algorithms.}
We present pseudo codes of the analytical reconstruction method (Section~\ref{s5ss2}) and optimization-based method (Section~\ref{s5ss3}) in Algorithms~\ref{alg:analytical} and \ref{alg:optimization}, respectively.
In Algorithm~\ref{alg:analytical}, we used noisy $\tilde{h}_\mathrm{blur}$ and $\tilde{e}$ instead of noiseless $h_\mathrm{blur}$ and $e$.
We estimated the oracle value $t_{\pi/2}$ and used the resulting value, denoted by $t_{\pi/2,\mathrm{anal}}$, as $t_{\pi/2}$.
In Algorithm~\ref{alg:optimization}, 
we implemented the gradient descent \eqref{eq:4_3_gd} leveraging the autograd functionality~\cite{Baydin:17} in PyTorch.
During the forward pass~(lines 5 and 6), 
non-differentiable operations were detached from the computational graph.
The initial values $h_{0,\mathrm{opt}}[0]$ and $t_{\pi/2,\mathrm{opt}}[0]$ were sampled from the normal distribution $\mathcal{N}(0,1)$ and the uniform distribution $\mathcal{U}(0,t_\mathrm{sens})$, respectively~(line 3).
We used the learning rates $\alpha_h=0.005$ and $\alpha_t=0.0005$.
The maximum number of iterations $I$ was $2500$.
In both Algorithms~\ref{alg:analytical} and \ref{alg:optimization},
we set $N=64$ and $C=0.15$.
All the experiments were conducted on the computing environment listed in Table~\ref{tab:spec_pc}.
Under these conditions,
the computation times of Algorithms~\ref{alg:analytical} and \ref{alg:optimization} were $0.04$ and $18.00$~s, respectively.
Thus, Algorithm~\ref{alg:analytical} is approximately $450$ times faster than Algorithm~\ref{alg:optimization}.

\section{Experiment}
\label{s6}
\subsection{Stability analysis}
\label{s6ss1}
To verify the stability of Algorithms~\ref{alg:analytical} and \ref{alg:optimization} with respect to hyperparameter selection, 
we evaluate their performance through computer simulations.
\vspace{2mm}

\noindent
\textbf{Dataset of object wavefront.}
We sampled $200$ images of size $346 \times 260$ from the CLIC2020 dataset~\cite{clic:21}.
The images were monochrome and had luminance values ranging from $0$ to $1$.
The pixel pitch was $18.5 \times 18.5$~\textmu m.
We generated $100$ complex object wavefronts $f$, 
where half of the sampled images were used as the intensity of $f$.
The remaining $100$ images were assigned as phase data, 
with the luminance values scaled by a factor of $2\pi$.
We placed $f$ at $z=120$~mm.
\vspace{2mm}

\begin{algorithm}[!t]
\caption{Analytical reconstruction for noisy data}
\label{alg:analytical}
\setstretch{1.2}
\footnotesize
\begin{algorithmic}[1]
\Require $\tilde{h}_\mathrm{blur}$ (noisy hologram), $\tilde{e}$ (noisy events), and $t_{\pi/2,\mathrm{anal}}$ (hyperparameter).
\State Compute $\tilde{E}_{0\rightarrow\pi/2}$ and $\tilde{E}_{0\rightarrow\pi}$.
\Comment{Use \eqref{eq:4_2_acc}, $t=t_{\pi/2,\mathrm{anal}}$ for $\tilde{E}_{0\rightarrow\pi/2}$, $t=t_\mathrm{sens}$ for $\tilde{E}_{0\rightarrow\pi}$, and $e=\tilde{e}$.}
\State Compute $\tilde{\mathcal{E}}$.
\Comment{Use \eqref{eq:4_2_vare} and $e=\tilde{e}$.}
\State Compute $\tilde{g}$.
\Comment{Use \eqref{eq:4_2_g}, $E_{0\rightarrow\pi/2}=\tilde{E}_{0\rightarrow\pi/2}$, $E_{0\rightarrow\pi}=\tilde{E}_{0\rightarrow\pi}$, $\mathcal{E}=\tilde{\mathcal{E}}$, and $h_\mathrm{blur}=\tilde{h}_\mathrm{blur}$.}
\State Compute $\tilde{f}$.
\Comment{Use reverse Fresnel diffraction and $g=\tilde{g}$.}
\end{algorithmic}
\end{algorithm}

\begin{algorithm}[!t]
\caption{Optimization-based reconstruction using autograd in PyTorch}
\label{alg:optimization}
\setstretch{1.2}
\footnotesize
\begin{algorithmic}[1]
\Require $\tilde{h}_\mathrm{blur}$, $\tilde{e}$, and $\lambda$ (hyperparameter).
\State Compute $\tilde{\mathcal{E}}$.
\Comment{Use \eqref{eq:4_2_vare} and $e=\tilde{e}$.}
\State Compute $\tilde{E}_{0\rightarrow\pi}$.
\Comment{Use \eqref{eq:4_2_acc}, $t=t_\mathrm{sens}$, and $e=\tilde{e}$.}
\State Initialize $h_{0,\mathrm{opt}}[0]$ and $t_{\pi/2,\mathrm{opt}}[0]$.
\Comment{Drawn from $\mathcal{N}(0,1)$ and $\mathcal{U}(0,t_\mathrm{sens})$.}
\For{$i=0$ \textbf{to} $I$}
\State Compute $\tilde{E}_{0\rightarrow\phi(t_{\pi/2,\mathrm{opt}}[i])}$.
\Comment{Use \eqref{eq:4_2_acc}, $t=t_{\pi/2,\mathrm{opt}}[i]$, and $e=\tilde{e}$.}
\State Compute $L(h_{0,\mathrm{opt}}[i],t_{\pi/2,\mathrm{opt}}[i])$.
\Comment{Use \eqref{eq:loss}, $h_{0,\mathrm{opt}}=h_{0,\mathrm{opt}}[i]$, and $t_{\pi/2,\mathrm{opt}}=t_{\pi/2,\mathrm{opt}}[i]$.}
\State Compute $h_{0,\mathrm{opt}}[i+1]$ and $t_{\pi/2,\mathrm{opt}}[i+1]$ via backpropagation.
\Comment{Equivalent to \eqref{eq:4_3_gd}.}
\EndFor
\State Compute $\tilde{f}$.
\Comment{Use reverse Fresnel diffraction and $g=\tilde{g}({h_{0,\mathrm{opt}}[I],t_{\pi/2,\mathrm{opt}}[I]})$.}
\end{algorithmic}
\end{algorithm}

\begin{table*}[!t]
\centering
\renewcommand{\arraystretch}{1.4}
\caption{Computing environment.}
\label{tab:spec_pc}
\footnotesize
\begin{tabular}{cwc{58mm}}\bhline
CPU         & Intel Core i9-13900KF (24 cores, 32 threads) \\
Main memory & 64 GB \\
OS          & Ubuntu 20.04 LTS \\
Language \& framework & Python 3.11.13 \& PyTorch 2.7.1\\\bhline
\end{tabular}
\end{table*}

\noindent
\textbf{Dataset of blurred hologram and event.} 
We simulated $100$ noiseless blurred holograms $h_\mathrm{blur}$ from the above-mentioned object wavefronts, using $M=500$ and the front-parallel $r$ with a constant intensity of $0.5$.
We implemented the Fresnel diffraction integral using the band-limited angular spectrum method~\cite{Matsushima:09}.
The pixel pitch of the holograms was $18.5 \times 18.5$~\textmu m.
We computed noisy holograms $\tilde{h}_\mathrm{blur}$ by adding zero-mean Gaussian noise and applying 8-bit quantization to $h_\mathrm{blur}$;
the noise standard deviation $\sigma_\mathrm{frame}$ was set to $0$, $0.01$, or $0.02$.
Following \cite{Hu:21}, we assumed that the contrast threshold fluctuates pixel by pixel.
In the event generation process,
we used Gaussian noise of mean $0.15$ and standard deviation $\sigma_\mathrm{event}=0.015$ as pixel-wise contrast thresholds.
When computing \eqref{eq:4_2_evt_rel}, we applied the floor operation to its right-hand side.
\vspace{2mm}

\begin{figure}[!t]
\centering
\begin{subfigure}{0.999\textwidth}
\centering
\includegraphics[scale=1.1]{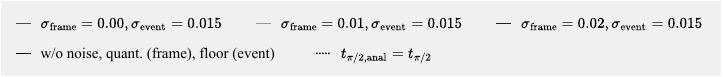}
\end{subfigure}
\begin{subfigure}{0.49\textwidth}
\centering
\includegraphics[width=0.9\linewidth]{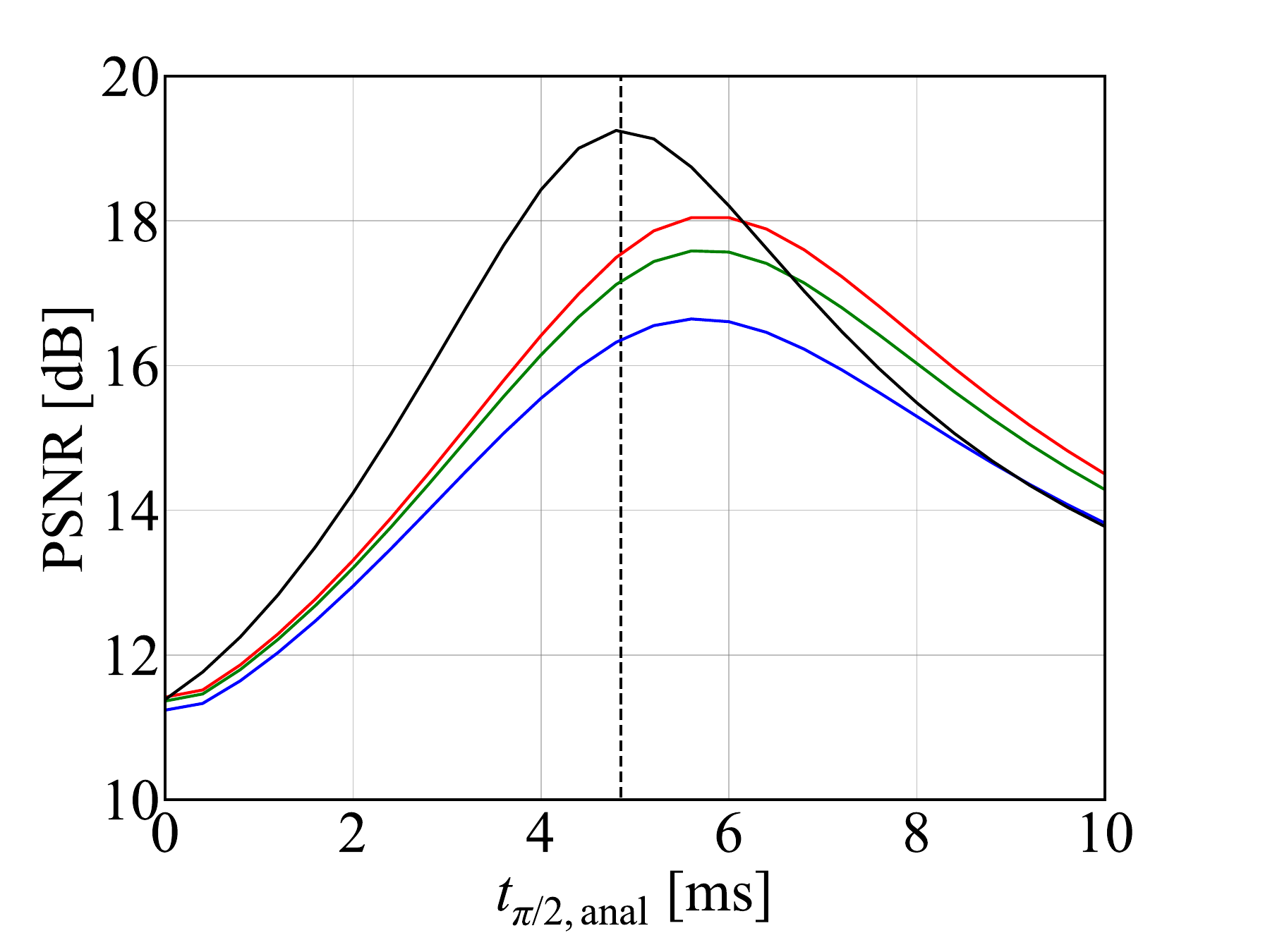}
\caption{Intensity of $\tilde{f}$}
\label{fig:ana_i}
\end{subfigure}
\begin{subfigure}{0.49\textwidth}
\centering
\includegraphics[width=0.9\linewidth]{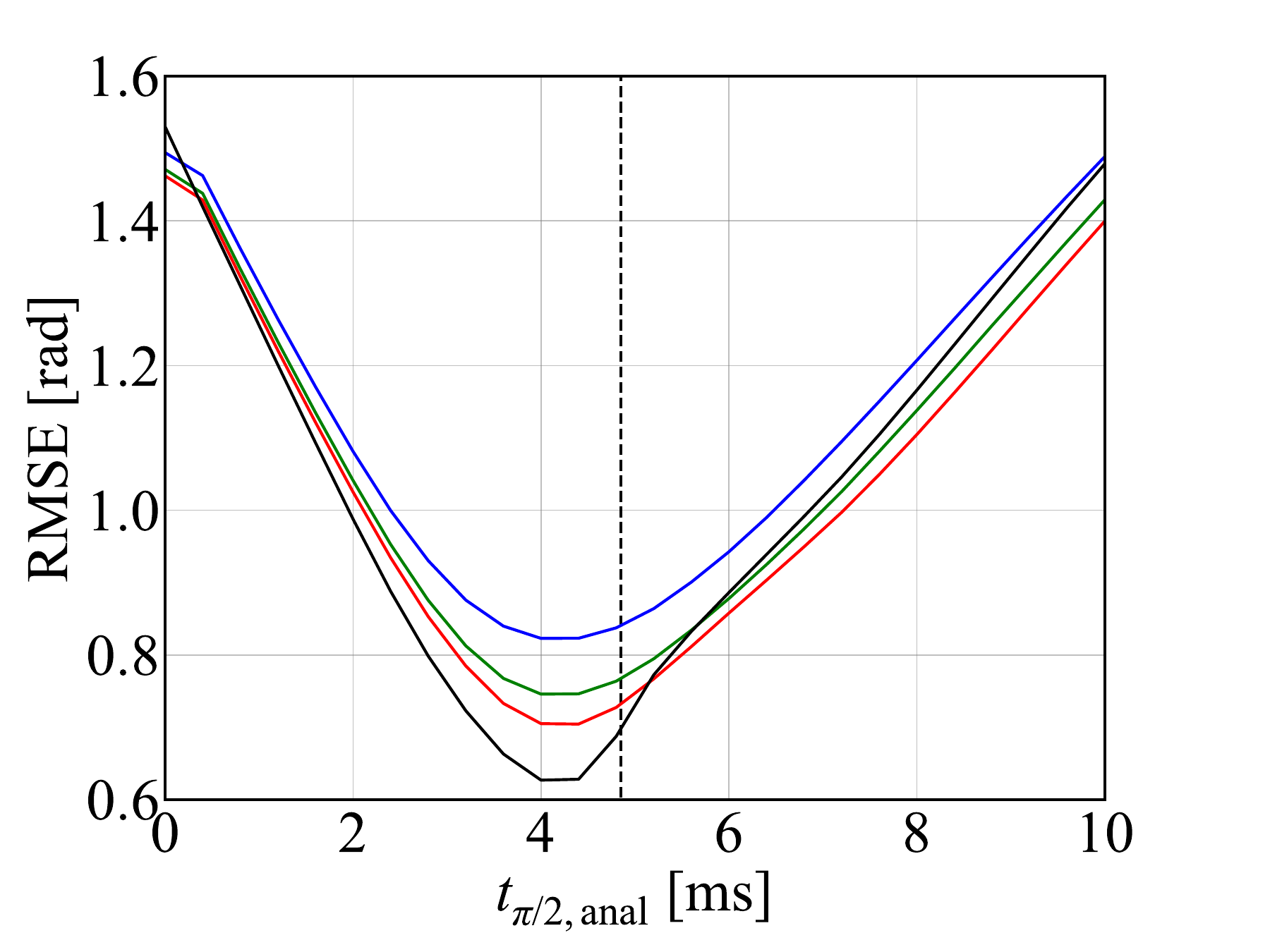}
\caption{Phase of $\tilde{f}$}
\label{fig:ana_p}
\end{subfigure}
\caption{Errors by Algorithm~\ref{alg:analytical} against $t_{\pi/2,\mathrm{anal}}$.}
\label{fig:ana}
\end{figure}

\begin{figure}[!t]
\centering
\begin{subfigure}{0.999\textwidth}
\centering
\includegraphics[scale=1.1]{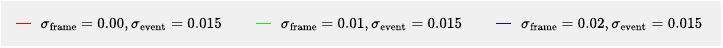}
\end{subfigure}
\begin{subfigure}{0.49\textwidth}
\centering
\includegraphics[width=0.9\linewidth]{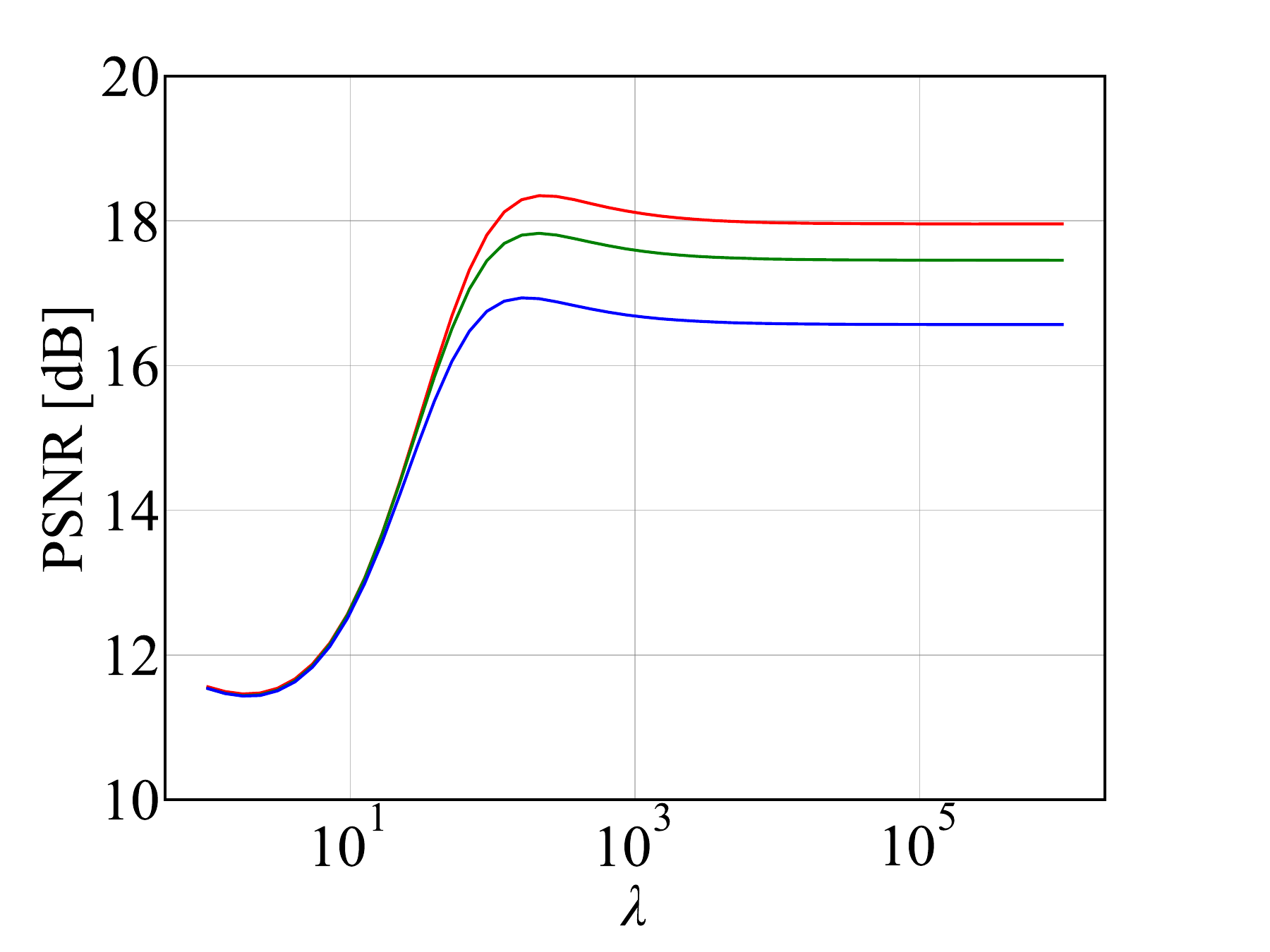}
\caption{Intensity of $\tilde{f}$}
\label{fig:opt_a}
\end{subfigure}
\begin{subfigure}{0.49\textwidth}
\centering
\includegraphics[width=0.9\linewidth]{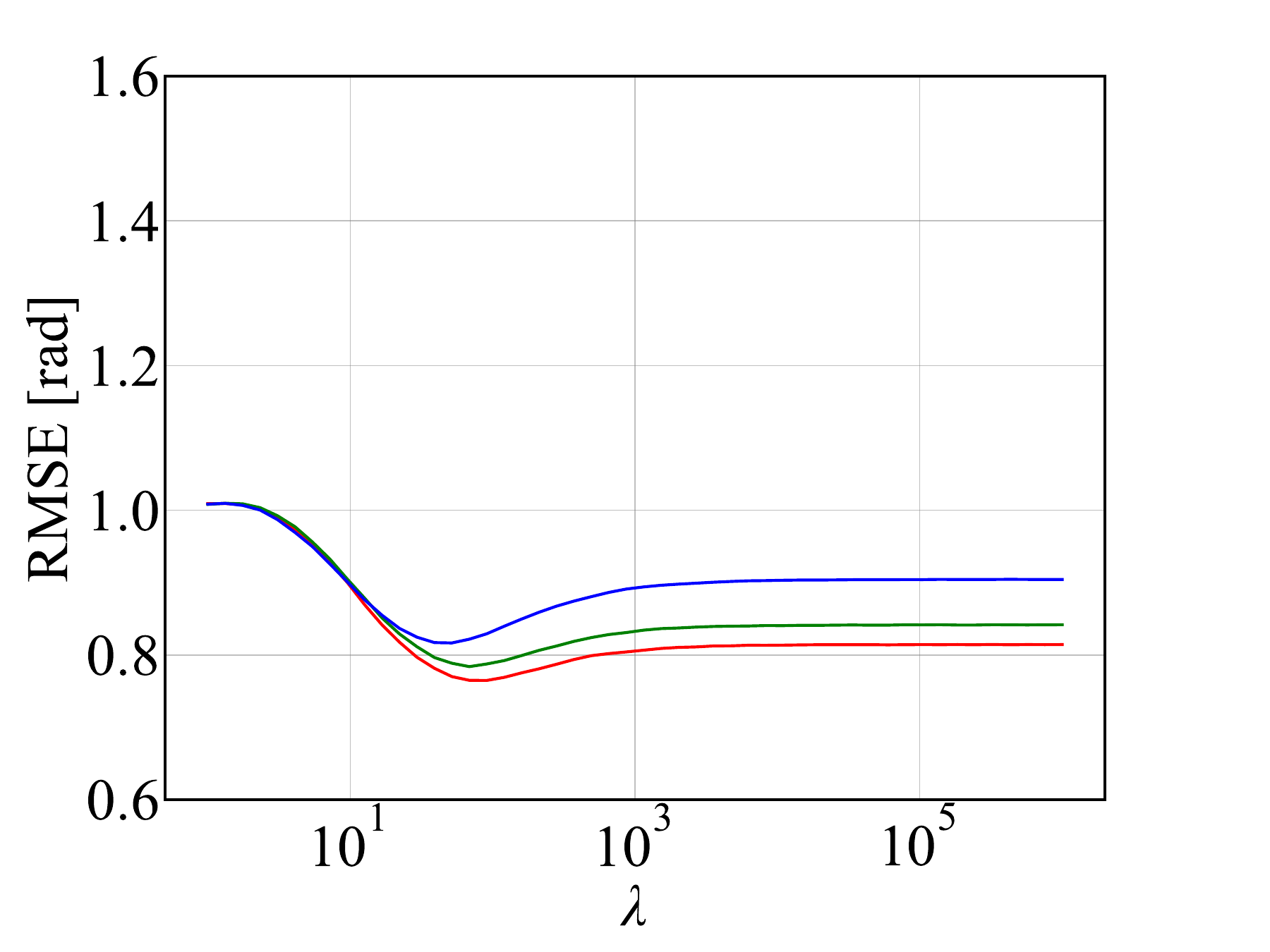}
\caption{Phase of $\tilde{f}$}
\label{fig:opt_b}
\end{subfigure}
\caption{Errors by Algorithm~\ref{alg:optimization} against $\lambda$.}
\label{fig:opt}
\end{figure}

\begin{figure}[!t]
\centering
\begin{subfigure}{0.999\textwidth}
\centering
\includegraphics[scale=1.1]{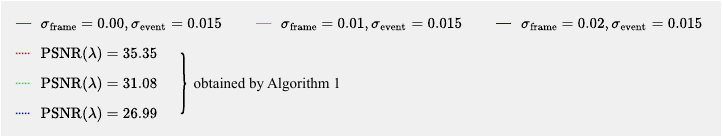}
\end{subfigure}
\begin{subfigure}{0.49\textwidth}
\centering
\includegraphics[width=0.9\linewidth]{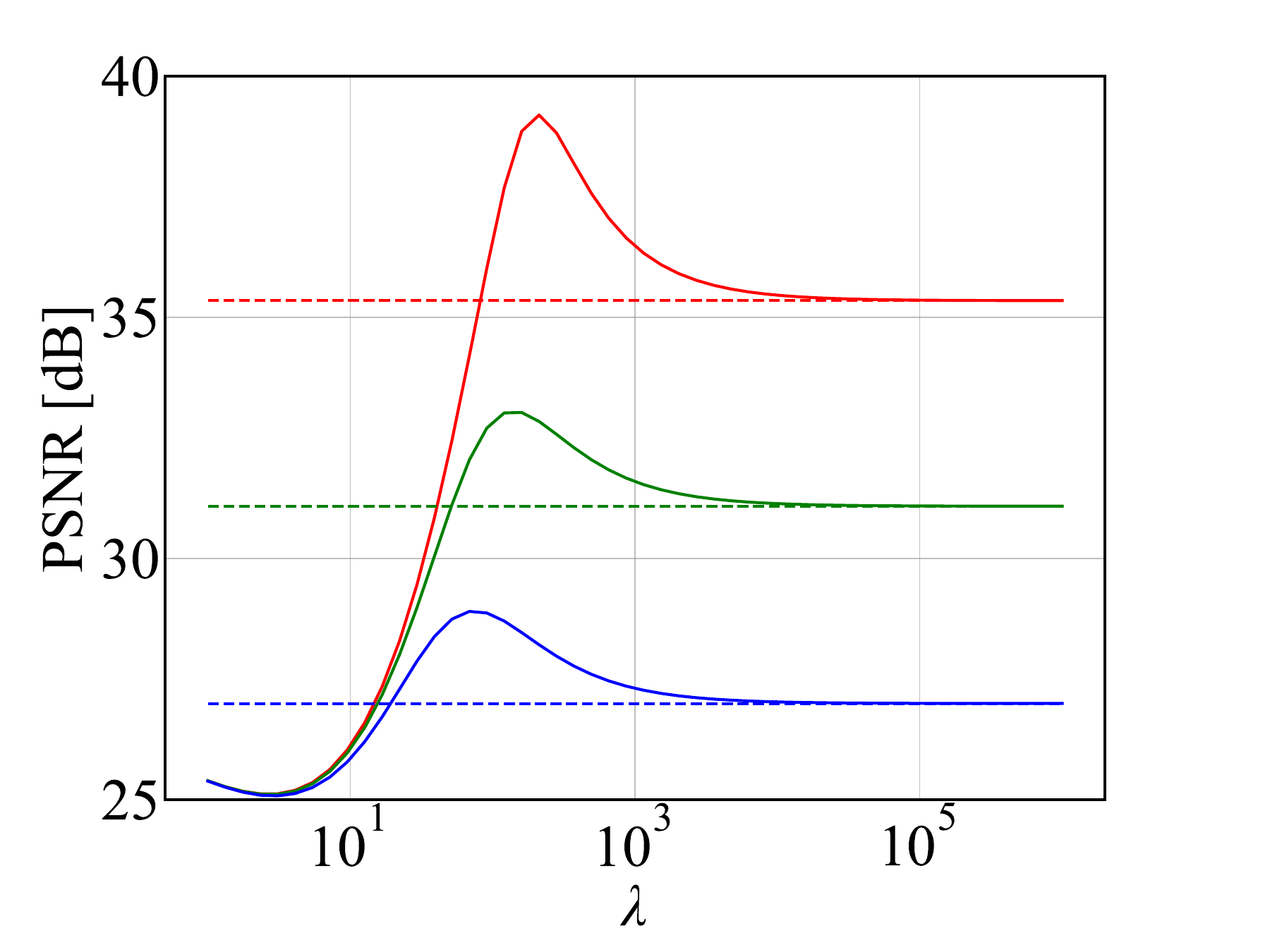}
\caption{$h_{0,\mathrm{opt}}[I]$}
\label{fig:opt_c}
\end{subfigure}
\begin{subfigure}{0.49\textwidth}
\centering
\includegraphics[width=0.9\linewidth]{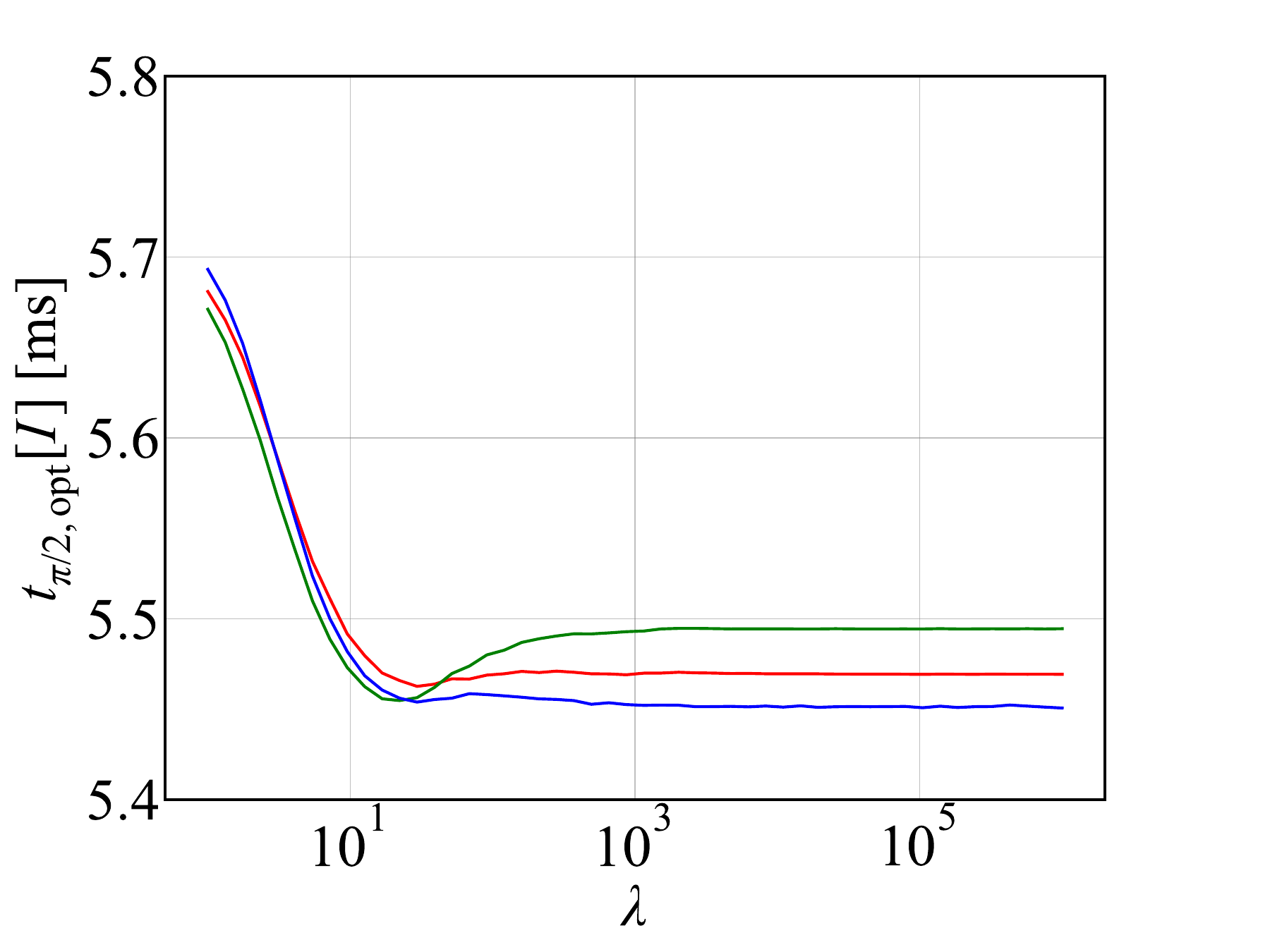}
\caption{$t_{\pi/2,\mathrm{opt}}[I]$}
\label{fig:opt_d}
\end{subfigure}
\caption{PSNR of $h_{0,\mathrm{opt}}[I]$ and reconstructed time $t_{\pi/2,\mathrm{opt}}[I]$.}
\label{fig:opt_psnr}
\end{figure}

\begin{figure}[!t]
\centering
\begin{subfigure}{0.999\textwidth}
\centering
\includegraphics[scale=1.1]{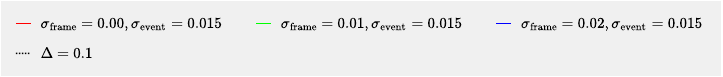}
\end{subfigure}
\begin{subfigure}{0.49\textwidth}
\centering
\includegraphics[width=0.9\linewidth]{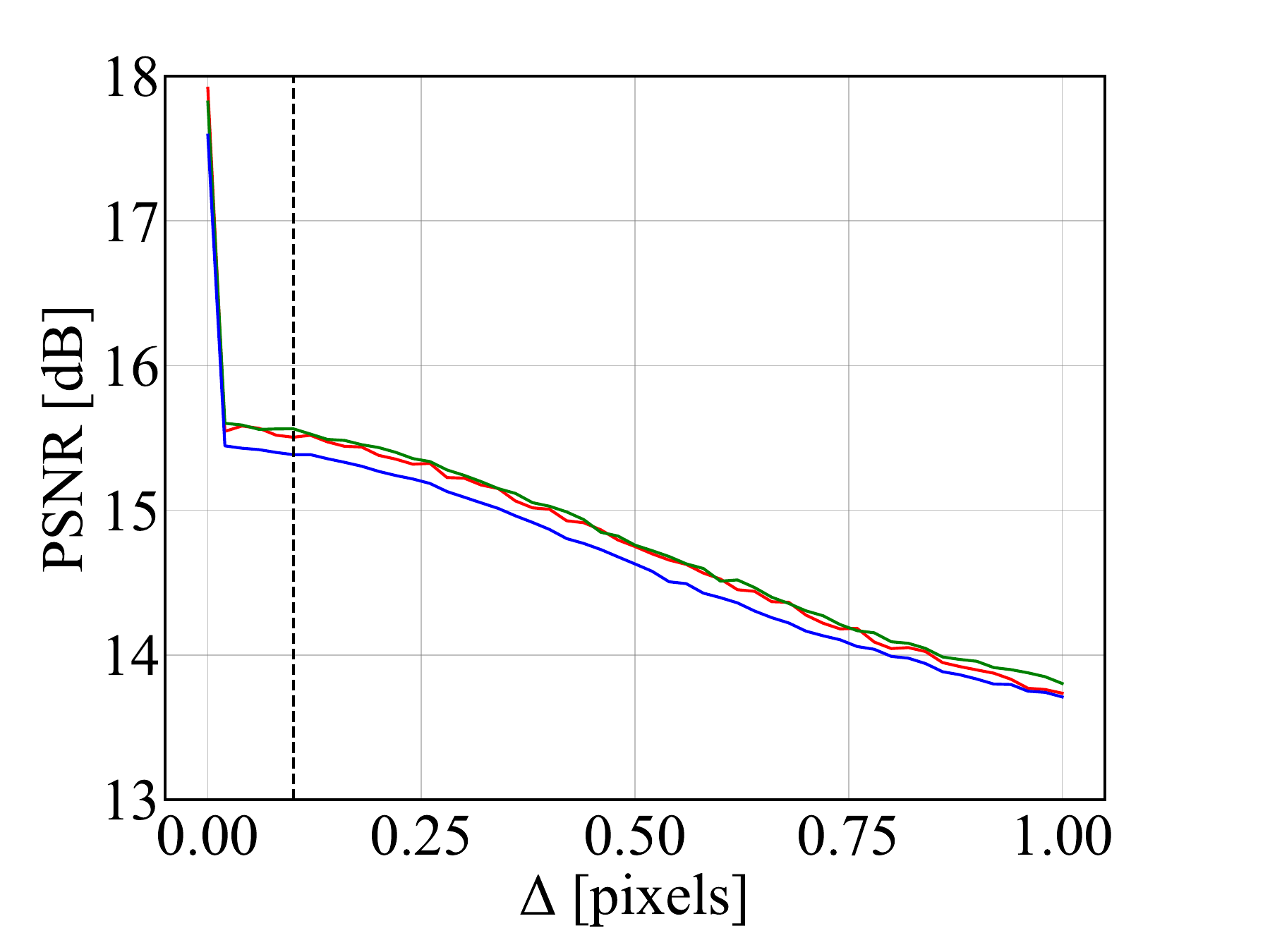}
\caption{Intensity of $\tilde{f}$}
\end{subfigure}
\begin{subfigure}{0.49\textwidth}
\centering
\includegraphics[width=0.9\linewidth]{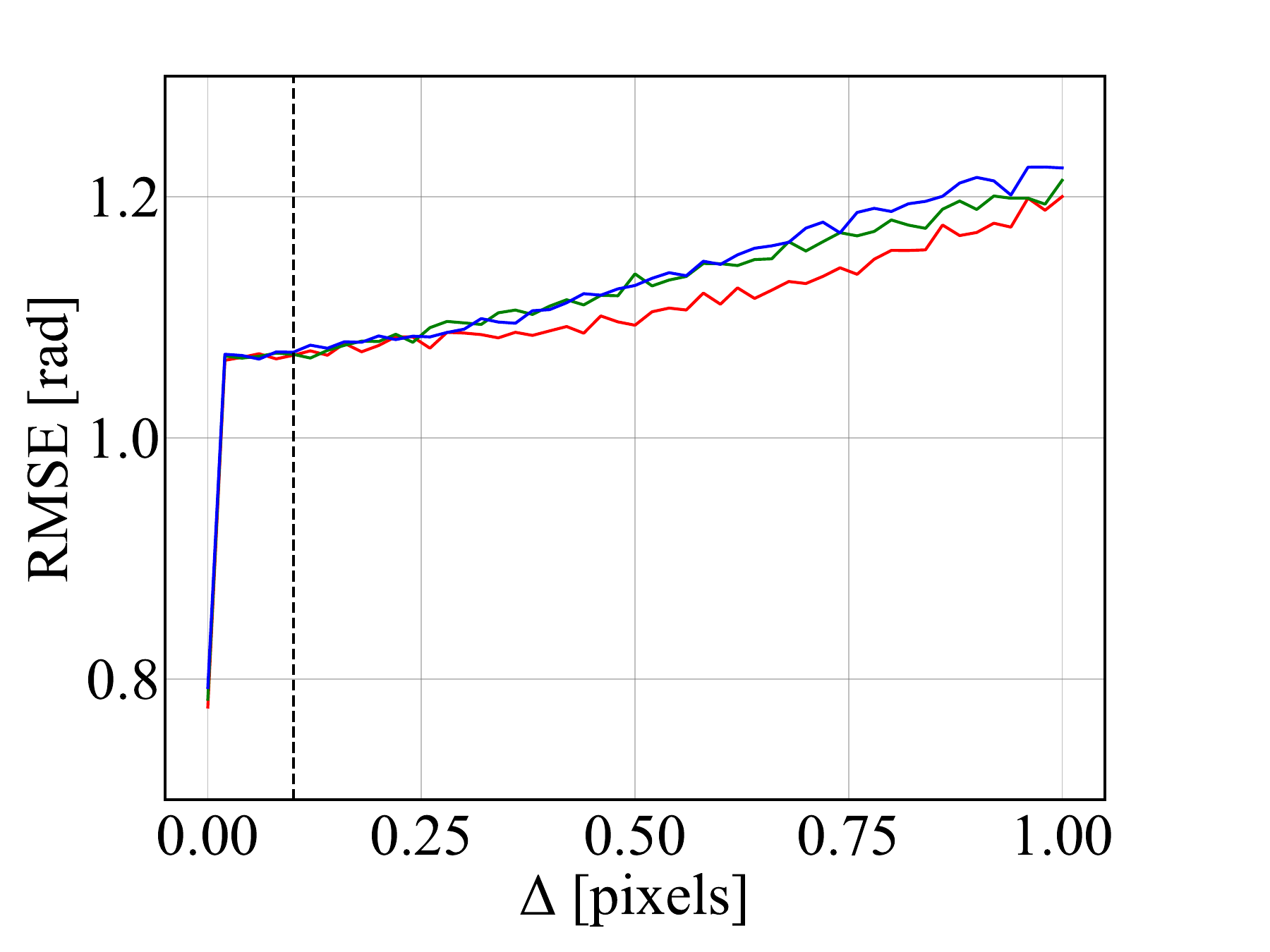}
\caption{Phase of $\tilde{f}$}
\end{subfigure}
\caption{Errors by Algorithm~\ref{alg:optimization}~($\lambda=500$) against motion $\Delta$.}
\label{fig:opt_appen}
\end{figure}

\noindent
\textbf{Response characteristic of SLM.}
Following \cite{Thalhammer:13}, we modeled the phase transition from $0$ to $\pi$ as the following non-linear time-evolution function:
\begin{equation}
\phi(t)=\pi\Big(1-\exp(-t/t_\mathrm{speed})\Big),
\label{eq:5_1_phase}
\end{equation}
where $t_\mathrm{speed}$ controls the transition speed and was set to $7$~ms.
Note that $t_{\pi/2}$ was $4.85$~ms.
\vspace{2mm}

\noindent
\textbf{Evaluation metrics.}
We measured peak signal-to-noise ratio (PSNR) values~[dB] between the intensities of $f$ and $\tilde{f}$.
The quality of the reconstructed phases was evaluated using root-mean-squared error (RMSE)~[rad] values,
where the phases corresponding to zero intensities were removed from the computation.
%
%
\vspace{2mm}

\noindent
\textbf{Remark~2.} 
The errors cannot be reduced to zero due to the band-limitation of the Fresnel diffraction integral.
\vspace{2mm}

\noindent
\textbf{Results on Algorithm~\ref{alg:analytical}.} 
Figure~\ref{fig:ana} shows the PSNR and RMSE values as functions of $t_{\pi/2,\mathrm{anal}}$.
In the noiseless scenario~(black lines),
the maximum and minimum values in \ref{fig:ana_i} and \ref{fig:ana_p}, respectively,
appeared near $t_{\pi/2,\mathrm{anal}} = t_{\pi/2}$.
However, in the noisy counterpart~(colored lines),
PSNR and RMSE values were optimized at $t_{\pi/2,\mathrm{anal}}$ that is apart from $t_{\pi/2}$.
Moreover, even a small deviation from the optimal $t_{\pi/2,\mathrm{anal}}$ value resulted in a sharp decline in the reconstruction quality.
These characteristics make the analytical method difficult to use for noisy real-world data as the determination of the optimal $t_{\pi/2,\mathrm{anal}}$ is non-trivial.
\vspace{2mm}

\noindent
\textbf{Remark 3.} For the red, green, and blue lines in Fig.~\ref{fig:ana}, PSNR values between ideal $h_0$ and $\tilde{h}_{0}$, i.e., $\tilde{h}_\mathrm{blur}/\tilde{\mathcal{E}}$, were $35.35$, $31.08$, and $26.99$~dBs, respectively.
\vspace{2mm}

\noindent
\textbf{Results on Algorithm~\ref{alg:optimization}.} 
Next, we demonstrate the stability of Algorithm~\ref{alg:optimization} by changing $\lambda$.
Figures~\ref{fig:opt_a} and \ref{fig:opt_b} summarize the results.
When $100 \leq \lambda$, Algorithm~\ref{alg:optimization} consistently achieved nearly optimal reconstruction.
This result indicates that, unlike Algorithm~\ref{alg:analytical},
Algorithm~\ref{alg:optimization} is largely insensitive to the choice of hyperparameter.
Moreover, Algorithm~\ref{alg:optimization} stably presents reasonable reconstruction quality by choosing a large $\lambda$.
\vspace{2mm}

\noindent
\textbf{Remark 4.} 
Figure~\ref{fig:opt_c} shows PSNR values between $h_0$ and reconstructed $h_{0,\mathrm{opt}}[I]$,
which were higher than those of Algorithm~\ref{alg:analytical}.
Figure~\ref{fig:opt_d} indicates $t_{\pi/2,\mathrm{opt}}[I]$,
which were slightly larger than $t_{\pi/2}$, similar to Fig.~\ref{fig:ana_i}.
\vspace{2mm}

\begin{figure}[!t]
\centering
\begin{subfigure}{1\textwidth}
\hspace{17.2mm}
\includegraphics[scale=1.1]{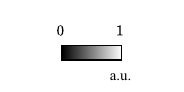}
\end{subfigure}\\
\begin{subfigure}{0.25\textwidth}
\centering
\includegraphics[width=0.9\linewidth]{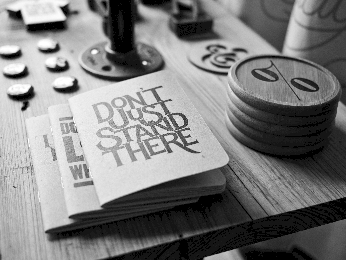}
\caption*{ }\vspace{2mm}
\includegraphics[width=0.9\linewidth]{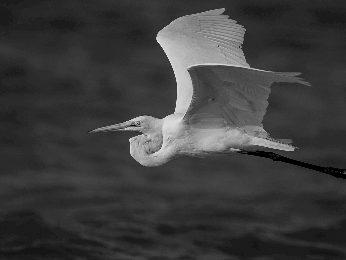}
\caption*{ }
\caption{Original}
\end{subfigure}\hspace{4mm}
%
\begin{subfigure}{0.25\textwidth}
\centering
\includegraphics[width=0.9\linewidth]{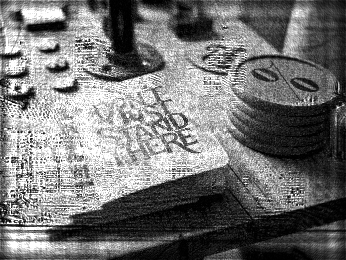}
\caption*{15.88}\vspace{2mm}
\includegraphics[width=0.9\linewidth]{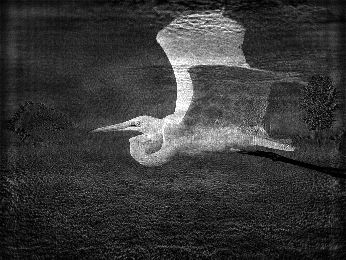}
\caption*{21.85}
\caption{Algorithm~\ref{alg:analytical}~($\Delta=0$)}
\end{subfigure}\hspace{4mm}
%
\begin{subfigure}{0.25\textwidth}
\centering
\includegraphics[width=0.9\linewidth]{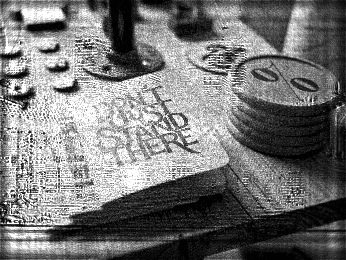}
\caption*{16.26}\vspace{2mm}
\includegraphics[width=0.9\linewidth]{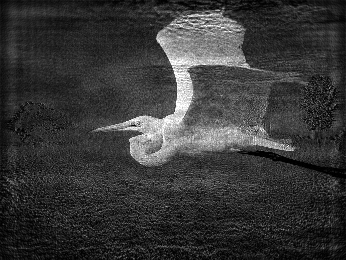}
\caption*{22.01}
\caption{Algorithm~\ref{alg:optimization}~($\Delta=0$)}
\label{fig:rec_mag_sim_static}
\end{subfigure}\vspace{4mm}\\
%
\begin{subfigure}{0.25\textwidth}
\centering
\includegraphics[width=0.9\linewidth]{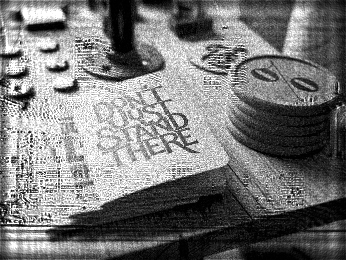}
\caption*{16.56}\vspace{2mm}
\includegraphics[width=0.9\linewidth]{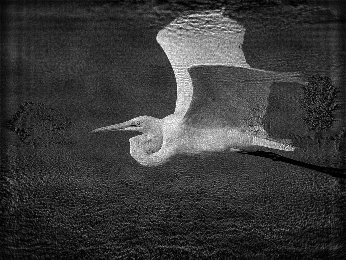}
\caption*{22.78}
\caption{F-PSDH-ME~($\Delta=0$)}
\end{subfigure}\hspace{4mm}
%
\begin{subfigure}{0.25\textwidth}
\centering
\includegraphics[width=0.9\linewidth]{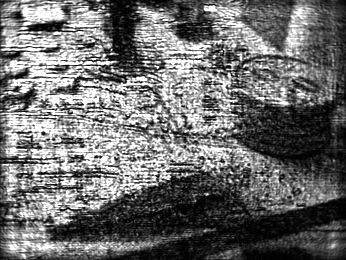}
\caption*{12.54}\vspace{2mm}
\includegraphics[width=0.9\linewidth]{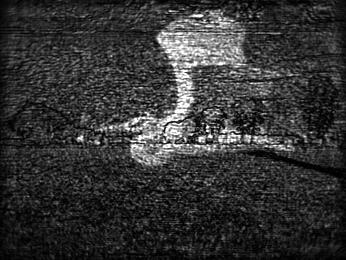}
\caption*{17.94}
\caption{F-PSDH-SE~($\Delta=0$)}
\label{fig:rec_mag_sim_psdh}
\end{subfigure}\hspace{4mm}
\begin{subfigure}{0.25\textwidth}
\centering
\includegraphics[width=0.9\linewidth]{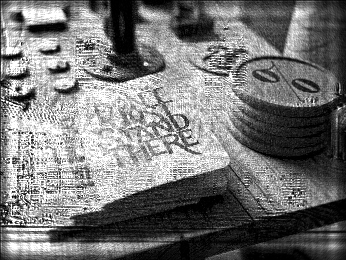}
\caption*{14.98}\vspace{2mm}
\includegraphics[width=0.9\linewidth]{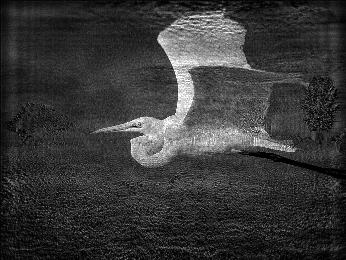}
\caption*{21.02}
\caption{Algorithm~\ref{alg:optimization}~($\Delta=0.1$)}
\label{fig:rec_mag_sim_dynamic}
\end{subfigure}
\caption{Reconstructed intensities with PSNRs~(higher is better)~[dB] in simulation. $\sigma_\mathrm{frame}=0.01$,  $\sigma_\mathrm{event}=0.015$, $t_{\pi/2,\mathrm{anal}}=4.85~\mathrm{ms}$ in Algorithm~\ref{alg:analytical}, and $\lambda=500$ in Algorithm~\ref{alg:optimization}.}
\vspace{4mm}
\label{fig:rec_mag_sim}
\end{figure}

\begin{figure}[!t]
\centering
\begin{subfigure}{1\textwidth}
\hspace{17.2mm}
\includegraphics[scale=1.1]{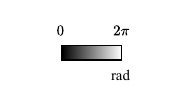}
\end{subfigure}\\
\hspace{0.0mm}
\begin{subfigure}{0.25\textwidth}
\centering
\includegraphics[width=0.9\linewidth]{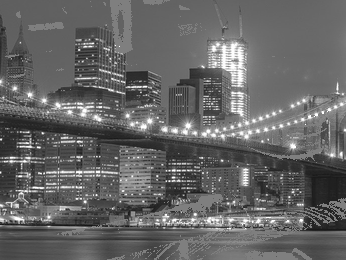}
\caption*{ }\vspace{2mm}
\includegraphics[width=0.9\linewidth]{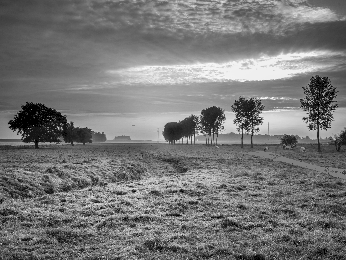}
\caption*{ }
\caption{Original}
\end{subfigure}\hspace{4mm}
%
\begin{subfigure}{0.25\textwidth}
\centering
\includegraphics[width=0.9\linewidth]{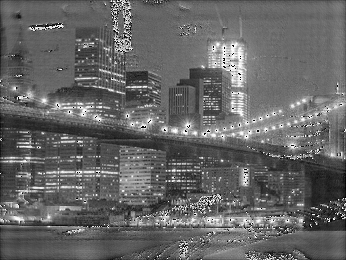}
\caption*{0.37}\vspace{2mm}
\includegraphics[width=0.9\linewidth]{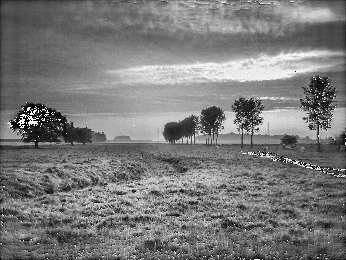}
\caption*{0.48}
\caption{Algorithm~\ref{alg:analytical}~($\Delta=0$)}
\end{subfigure}\hspace{4mm}
%
\begin{subfigure}{0.25\textwidth}
\centering
\includegraphics[width=0.9\linewidth]{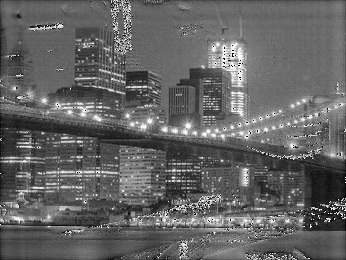}
\caption*{0.32}\vspace{2mm}
\includegraphics[width=0.9\linewidth]{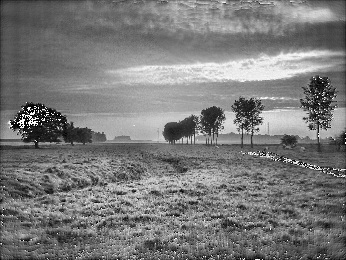}
\caption*{0.47}
\caption{Algorithm~\ref{alg:optimization}~($\Delta=0$)}
\label{fig:rec_phase_sim_static}
\end{subfigure}
\vspace{4mm}\\
%
\begin{subfigure}{0.25\textwidth}
\centering
\includegraphics[width=0.9\linewidth]{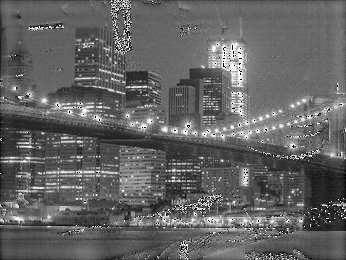}
\caption*{0.36}\vspace{2mm}
\includegraphics[width=0.9\linewidth]{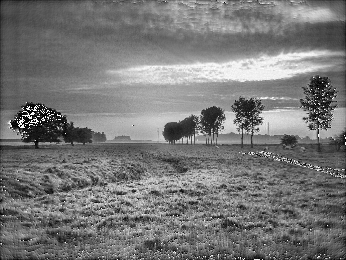}
\caption*{0.47}
\caption{F-PSDH-ME~($\Delta=0$)}
\end{subfigure}\hspace{4mm}
%
\begin{subfigure}{0.25\textwidth}
\centering
\includegraphics[width=0.9\linewidth]{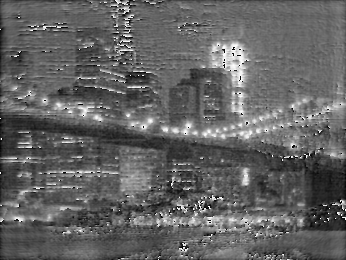}
\caption*{0.57}\vspace{2mm}
\includegraphics[width=0.9\linewidth]{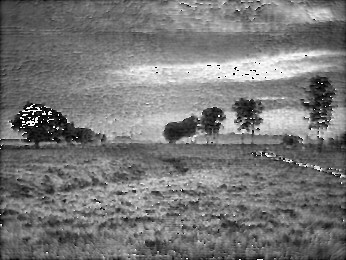}
\caption*{0.68}
\caption{F-PSDH-SE~($\Delta=0$)}
\label{fig:rec_phase_sim_psdh}
\end{subfigure}\hspace{4mm}
\begin{subfigure}{0.25\textwidth}
\centering
\includegraphics[width=0.9\linewidth]{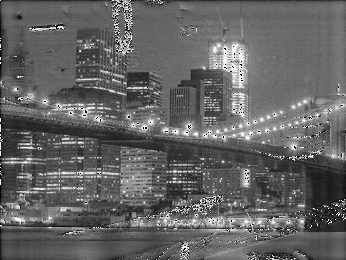}
\caption*{0.76}\vspace{2mm}
\includegraphics[width=0.9\linewidth]{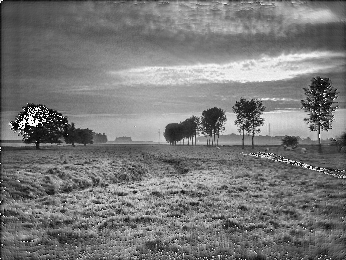}
\caption*{0.88}
\caption{Algorithm~\ref{alg:optimization}~($\Delta=0.1$)}
\label{fig:rec_phase_sim_dynamic}
\end{subfigure}
\caption{Reconstructed phases with RMSEs~(lower is better)~[rad] in simulation. We used same parameters as in Fig.~\ref{fig:rec_mag_sim}.}
\label{fig:rec_phase_sim}
\end{figure}

\noindent
\textbf{Dynamic scenario.} 
To evaluate the stability of Algorithm~\ref{alg:optimization} for dynamic objects,
the object wavefronts were shifted toward the upper right at a constant velocity during the exposure.
The total shift over the exposure ranged from $0$ to $1$ pixels.
Figure~\ref{fig:opt_appen} presents the results, 
where $\lambda$ was fixed to $500$ and $\Delta$ denotes the shift value.
When $0.1<\Delta$, the errors gradually increased.
In contrast, for $0 < \Delta \leq 0.1$,
Algorithm~\ref{alg:optimization} consistently maintained errors at a low level.
These results indicate that Algorithm~\ref{alg:optimization} is robust to small object motions.
\vspace{2mm}

\subsection{Comparison with F-PSDH}
\label{s6ss2}
To demonstrate the effectiveness of FE-PSDH,
we visualize the reconstructed wavefronts and compare them with those obtained by F-PSDH.
The experiments were conducted both on computer simulations and the optical system in Fig.~\ref{fig:impl}.
\vspace{2mm}

\noindent
\textbf{Configuration of F-PSDH.}
For the simulation,
we generated the noiseless $h_{0}$, $h_{\pi/2}$, and $h_{\pi}$ by varying $\phi$ in \eqref{eq:3_holo},
where the intensity and phase of $f$ were produced in the same manner as in Section~\ref{s6ss1}.
We added Gaussian noise of variance $\sigma_\mathrm{frame}=0.01$ and applied 8-bit quantization to the generated holograms.
For the optical experiment,
we recorded the holograms by the image sensor in the hybrid EVS, DAVIS346, 
through the three exposures and two phase shifts.
We employed the following two-types of F-PSDH~\cite{Yamaguchi:97,Awatsuji:06}:
\begin{itemize}
\item \textit{F-PSDH-ME~(multi exposure)}~\cite{Yamaguchi:97} records $h_0$, $h_{\pi/2}$, and $h_\pi$ with three individual exposures.
The object wavefronts were reconstructed by \eqref{eq:3_psdh} and the reverse Fresnel diffraction, where noisy holograms were used as $h_0$, $h_{\pi/2}$, and $h_{\pi}$.
\item \textit{F-PSDH-SE~(single exposure)}~\cite{Awatsuji:06} multiplexes $h_0$, $h_{\pi/2}$, and $h_{\pi}$ in a single frame, 
enabling the acquisition within a single exposure.
To simulate this multiplexing process, 
we reduced the resolutions of $h_0$, $h_{\pi/2}$, and $h_\pi$ by half.
The holograms were then refined by pseudo super-resolution and interpolation, see \cite{Awatsuji:06} for more details.
The object wavefronts were reconstructed in the same manner as in F-PSDH-ME.
\end{itemize}

\noindent
\textbf{Results on simulation.}
Figure~\ref{fig:rec_mag_sim} presents the reconstructed intensities and Fig.~\ref{fig:rec_phase_sim} shows their corresponding phases.
\begin{itemize} 
\item
Static scenario~($\Delta=0$):
Algorithm~\ref{alg:analytical} produced noisy results due to the misalignment of the empirical $t_{\pi/2,\mathrm{anal}}$ value and the optimal one.
In contrast, Algorithm~\ref{alg:optimization} exhibited results comparable to those by F-PSDH-ME.
Moreover, Algorithm~\ref{alg:optimization} outperformed F-PSDH-SE in terms of PSNR and RMSE values.
\item
Dynamic scenario~($\Delta=0.1$):
The PSNR and RMSE values of Algorithm~\ref{alg:optimization} (Figs.~\ref{fig:rec_mag_sim_dynamic} and \ref{fig:rec_phase_sim_dynamic}) were slightly degraded compared to those in the static case (Figs.~\ref{fig:rec_mag_sim_static} and \ref{fig:rec_phase_sim_static}).
However, even in the presence of object motion, the PSNR and RMSE values were comparable to, or better than, those of F-PSDH-SE in the static case (Figs.~\ref{fig:rec_mag_sim_psdh} and \ref{fig:rec_phase_sim_psdh}).
\end{itemize}

\vspace{2mm}

\begin{figure}[!t]
\centering
\begin{subfigure}{1\textwidth}
\hspace{17.2mm}
\includegraphics[scale=1.1]{images/bar1.pdf}
\end{subfigure}\\
\begin{subfigure}{0.25\textwidth}
\centering
\includegraphics[width=0.9\linewidth]{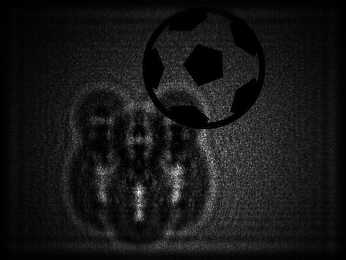}
\caption*{ }\vspace{2mm}
\includegraphics[width=0.9\linewidth]{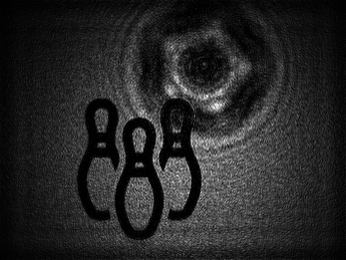}
\caption*{ }
\caption{Original (simulated)}
\label{fig:rec_real_org}
\end{subfigure}\hspace{4mm}
%
\begin{subfigure}{0.25\textwidth}
\centering
\includegraphics[width=0.9\linewidth]{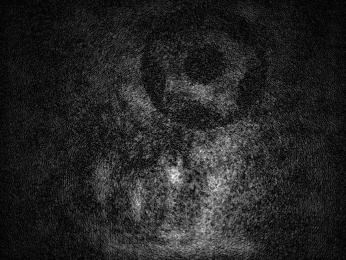}
\caption*{0.57}\vspace{2mm}
\includegraphics[width=0.9\linewidth]{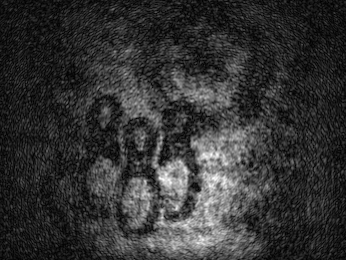}
\caption*{0.51}
\caption{Algorithm~\ref{alg:analytical}~(static)}
\label{fig:rec_real_b}
\end{subfigure}\hspace{4mm}
%
\begin{subfigure}{0.25\textwidth}
\centering
\includegraphics[width=0.9\linewidth]{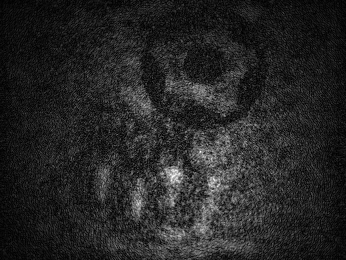}
\caption*{0.56}\vspace{2mm}
\includegraphics[width=0.9\linewidth]{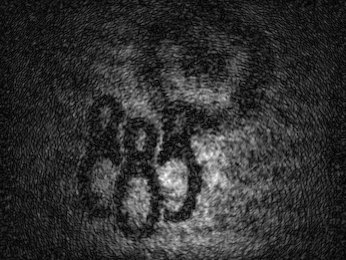}
\caption*{0.49}
\caption{Algorithm~\ref{alg:optimization}~(static)}
\end{subfigure}\vspace{4mm}\\
%
\begin{subfigure}{0.25\textwidth}
\centering
\includegraphics[width=0.9\linewidth]{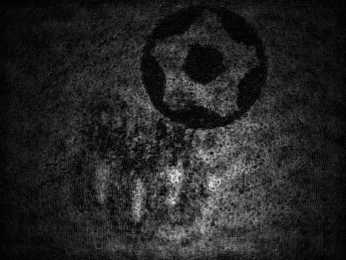}
\caption*{0.52}\vspace{2mm}
\includegraphics[width=0.9\linewidth]{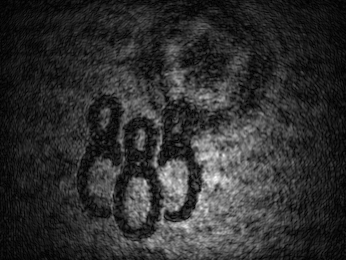}
\caption*{0.47}
\caption{F-PSDH-ME~(static)}
\end{subfigure}\hspace{4mm}
%
\begin{subfigure}{0.25\textwidth}
\centering
\includegraphics[width=0.9\linewidth]{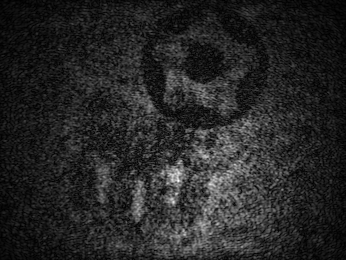}
\caption*{0.56}\vspace{2mm}
\includegraphics[width=0.9\linewidth]{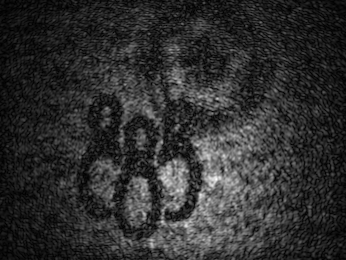}
\caption*{0.53}
\caption{F-PSDH-SE~(static)}
\label{fig:rec_real_e}
\end{subfigure}\hspace{4mm}
%
\begin{subfigure}{0.25\textwidth}
\centering
\includegraphics[width=0.9\linewidth]{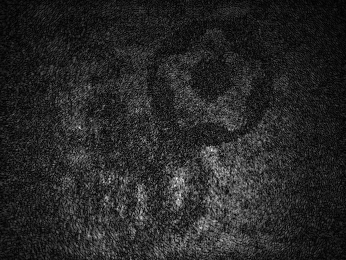}
\caption*{0.60}\vspace{2mm}
\includegraphics[width=0.9\linewidth]{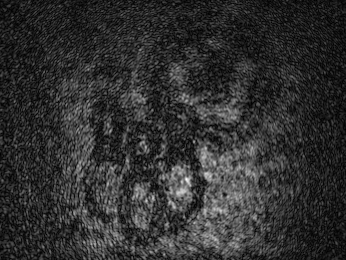}
\caption*{0.54}
\caption{Algorithm~\ref{alg:optimization}~(dynamic)}
\label{fig:rec_real_f}
\end{subfigure}
\caption{Reconstructed intensities with real-world data recorded by optical system in Fig.~\ref{fig:impl}. Values: LPIPS~(lower is better). Top: front focus~($z=120$~mm). Bottom: back focus~($z=270$~mm).}
\label{fig:rec_real}
\end{figure}

\noindent
\textbf{Results on optical experiment.}
Figures~\ref{fig:rec_real_b}--\ref{fig:rec_real_e} illustrate the reconstructed intensities using the real data for static objects, 
where all the methods can successfully reproduce the bokeh effect.
Similar to Fig.~\ref{fig:rec_mag_sim},
F-PSDH-ME and Algorithm~\ref{alg:optimization} presented fine details of the \textit{ball} and \textit{pins} at $z=120$ and $270$~mm, respectively.
To quantitatively evaluate the results,
we measured learned perceptual image patch similarity~(LPIPS; lower is better) values between the ideal intensity image (Fig.~\ref{fig:rec_real_org}) and the reconstructed ones~(Figs.~\ref{fig:rec_real_b}--\ref{fig:rec_real_e}).
The LPIPS value obtained by Algorithm~\ref{alg:optimization} was lower than that of Algorithm~\ref{alg:analytical}, indicating the effectiveness of Algorithm~\ref{alg:optimization}.
In addition,
Algorithm~\ref{alg:optimization} achieved comparable or lower LPIPS values than those by the F-PSDH methods.
\vspace{2mm}

\begin{figure}[!t]
\centering
\begin{subfigure}{0.25\textwidth}
\centering
\includegraphics[width=0.9\linewidth]{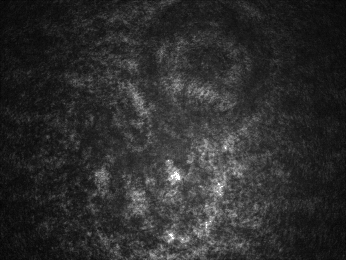}
\caption{Recorded frame}
\label{fig:obs_a}
\end{subfigure}\hspace{4mm}
%
\begin{subfigure}{0.25\textwidth}
\centering
\includegraphics[width=0.9\linewidth]{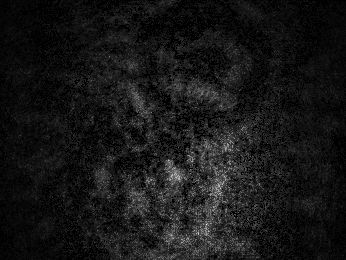}
\caption{Amplitude (Algo.~\ref{alg:analytical})}
\label{fig:obs_b}
\end{subfigure}\hspace{4mm}
%
\begin{subfigure}{0.25\textwidth}
\centering
\includegraphics[width=0.9\linewidth]{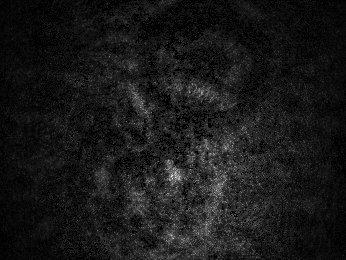}
\caption{Amplitude (Algo.~\ref{alg:optimization})}
\label{fig:obs_d}
\end{subfigure}
\vspace{4mm}\\
%
\begin{subfigure}{0.25\textwidth}
\centering
\includegraphics[scale=1.1]{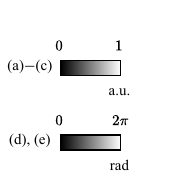}
\vspace{-3mm}
\caption*{ }
\end{subfigure}\hspace{3mm}
%
\begin{subfigure}{0.25\textwidth}
\centering
\includegraphics[width=0.9\linewidth]{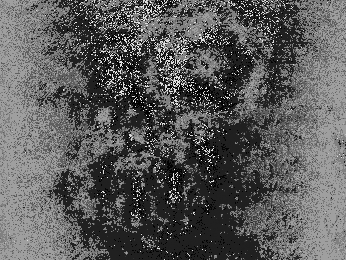}
\caption{Phase (Algo.~\ref{alg:analytical})}
\label{fig:obs_c}
\end{subfigure}\hspace{4mm}
%
\begin{subfigure}{0.25\textwidth}
\centering
\includegraphics[width=0.9\linewidth]{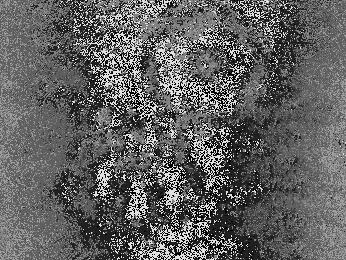}
\caption{Phase (Algo.~\ref{alg:optimization})}
\label{fig:obs_e}
\end{subfigure}
\caption{Recorded read-world intensity frame $\tilde{h}_\mathrm{blur}$ and reconstructed full-complex~(amplitude and phase) wavefronts $\tilde{g}$ at sensor plane.}
\label{fig:obs}
\end{figure}

\noindent
\textbf{Remark 5.} Figure~\ref{fig:obs_a} illustrates the intensity image $\tilde{h}_\mathrm{blur}$ recorded by our method.
Figures~\ref{fig:obs_b}--\ref{fig:obs_e} are amplitude and phase distributions of $\tilde{g}$ reconstructed by our method.
\vspace{2mm}

\noindent
\textbf{Dynamic scenario.}
To evaluate the stability of Algorithm~\ref{alg:optimization} for dynamic objects,
an OHP sheet printed with a ball pattern was slightly shaken during the exposure.
Figure~\ref{fig:rec_real_f} shows the reconstructed intensities of dynamic objects.
Although somewhat degraded, the target objects are clearly identified.
The LPIPS values were comparable to those without the motion.
This result suggests the potential applicability of the proposed method to dynamic holographic imaging scenarios, 
including digital holographic microscopy for observing dynamic cellular behavior~\cite{Kim:24}.
\vspace{2mm}

\noindent
Overall results demonstrate that FE-PSDH, particularly, Algorithm~\ref{alg:optimization}, 
can stably and reliably reconstruct object wavefronts while accelerating the acquisition time.

\section{Conclusion}
\label{s7}
Inspired by F-PSDH,
we introduced a novel wavefront reconstruction pipeline using the hybrid EVS.
The data for reconstructing wavefonts can be measured within the single exposure time alone,
which is the remarkable characteristic of our method.
We demonstrated the robustness of our method via simulation and optical experiments.
We believe our approach will potentially offer a new perspective on holography.

\bibliographystyle{IEEEtran}
\bibliography{p}

@article{Gigan:22,
author      = {S. Gigan and others},
title       = {Roadmap on wavefront shaping and deep imaging in complex media},
journal     = {Journal of Physics: Photonics},
volume      = {4},
number      = {4},
pages       = {042501},
year        = {2022}
}

@article{Goodman:67,
author      = {J. Goodman and R. W. Lawrence},
title       = {Digital image formation from electronically detected holograms},
journal     = {Applied Physics Letters},
volume      = {11},
number      = {3},
pages       = {77--79},
year        = {1967}
}

@article{Fienup:82,
author      = {J. R. Fienup},
title       = {Phase retrieval algorithms: a comparison},
journal     = {Applied Optics},
volume      = {21},
number      = {15},
pages       = {2758--2769},
year        = {1982}
}

@article{Streibl:84,
author      = {N. Streibl},
title       = {Phase imaging by the transport equation of intensity},
journal     = {Optics Communications},
volume      = {49},
number      = {1},
pages       = {6--10},
year        = {1984}
}

@article{Platt:01,
author      = {B. C. Platt and R. Shack},
title       = {History and principles of {Shack-Hartmann} wavefront sensing},
journal     = {Journal of Refractive Surgery},
volume      = {17},
number      = {5},
pages       = {S573--S577},
year        = {2001}
}

@article{Stoykova:14,
author      = {E. Stoykova and H. Kang and J. Park},
title       = {Twin-image problem in digital holography---a survey},
journal     = {Chinese Optics Letters},
volume      = {12},
number      = {6},
pages       = {060013},
year        = {2014}
}

@book{Goodman05,
author      = {J. W. Goodman},
title       = {Introduction to Fourier Optics},
publisher   = {Roberts \& Co. Publishers},
year        = {2005}
}

@article{Yamaguchi:97,
author      = {I. Yamaguchi and T. Zhang},
title       = {Phase-shifting digital holography},
journal     = {Optics Express},
volume      = {22},
number      = {16},
pages       = {1268--1270},
year        = {1997}
}

@article{Awatsuji:04,
author      = {Y. Awatsuji and M. Sasada and T. Kubota},
title       = {Parallel quasi-phase-shifting digital holography},
journal     = {Applied Phisics Letters},
volume      = {85},
number      = {6},
pages       = {1069--1071},
year        = {2004}
}

@article{Tahara:10,
author      = {T. Tahara and others},
title       = {Experimental demonstration of parallel two-step phase-shifting digital holography},
journal     = {Optics Express},
volume      = {18},
number      = {18},
pages       = {18975-18980},
year        = {2010}
}

@inproceedings{Habuchi:24,
author      = {S. Habuchi and K. Takahashi and C. Tsutake and T. Fujii and H. Nagahara},
title       = {Time-efficient light-field acquisition using coded aperture and events},
booktitle   = {Proceedings of the IEEE/CVF Conference on Computer Vision and Pattern Recognition},
pages       = {24923--24933},
year        = {2024}
}

@article{Lichtsteiner:08,
author      = {P. Lichtsteiner and C. Posch and T. Delbruck},
title       = {A {$128\times 128$} {$120$} {dB} 15 \textmu s latency asynchronous temporal contrast vision sensor},
journal     = {IEEE Journal of Solid-State Circuits},
volume      = {43},
number      = {2},
pages       = {566-576},
year        = {2008},
}

@inproceedings{Hu:21,
author      = {Y. Hu and S. C. Liu and T. Delbruck},
title       = {v2e: From video frames to realistic {DVS} events},
booktitle   = {Proceedings of IEEE/CVF Conference on Computer Vision and Pattern Recognition Workshops},
year        = {2021}
}

@inproceedings{Yu:24,
author      = {B. Yu and J. Ren and J. Han and F. Wang and J. Liang and B. Shi},
title       = {{EventPS}: Real-time photometric stereo using an event camera},
booktitle   = {Proceedings of IEEE/CVF Conference on Computer Vision and Pattern Recognition},
pages       = {9602-9611},
year        = {2024}
}

@article{Fu:23,
author      = {J. Fu and Y. Zhang and Y. Li and J. Li and Z. Xiong},
title       = {Fast {3D} reconstruction via event-based structured light with spatio-temporal coding},
journal     = {Light: Science \& Applications},
volume      = {31},
number      = {26},
pages       = {44588-44602},
year        = {2023}
}

@inproceedings{Uchiyama:24,
author      = {I. Uchiyama and C. Tsutake and K. Takahashi and T. Fujii},
title       = {Stabilizing digital holography using events},
booktitle   = {Proceedings of the 31st International Display Workshops},
volume      = {31},
year        = {2024}
}

@inproceedings{Wang:25,
author      = {C. Wang and S. Li and S. Zhu and E. Y. Lam},
title       = {Motion-resolved event-based holography},
booktitle   = {Proceedings of SPIE 13717, Advanced Optical Imaging Technologies VIII,},
year        = {2025}
}

@article{Ge:25,
author      = {Z. Ge and C. Wang and J. Huang and E. Y. Lam},
title       = {Event-driven neuromorphic holography for dynamic particle imaging},
journal     = {Optics Letters},
volume      = {50},
number      = {5},
pages       = {1496-1499},
year        = {2025}
}

@article{Baydin:17,
author      = {A. G. Baydin and B. A. Pearlmutter and A. A. Radul and J. M. Siskind},
title       = {Automatic differentiation in machine learning: A survey},
journal     = {Journal of Machine Learning Research},
volume      = {18},
number      = {1},
pages       = {5595-5637},
year        = {2017},
}

@article{Thalhammer:13,
author      = {G. Thalhammer and R. W. Bowman and G. D. Love and M. J. Padgett and M. R.-Marte},
title       = {Speeding up liquid crystal {SLMs} using overdrive with phase change reduction},
journal     = {Journal of Machine Learning Research},
volume      = {21},
number      = {2},
pages       = {1779-1797},
year        = {2013},
}

@article{Matsushima:09,
author      = {K. Matsushima and T. Shimobaba},
title       = {Band-limited angular spectrum method for numerical simulation of free-space propagation in far and near fields},
journal     = {Optics Express},
volume      = {17},
number      = {22},
pages       = {19662-19673},
year        = {2009},
}

@article{Awatsuji:06,
author      = {Y. Awatsuji and M. Sasada and A. Fujii and T. Kubota
},
title       = {Scheme to improve the reconstructed image in parallel quasi-phase-shifting digital holography},
journal     = {Applied Optics},
volume      = {45},
number      = {5},
pages       = {968-974},
year        = {2006},
}

@article{Kim:24,
author      = {J. Kim  and S. J. Lee},
title       = {Digital in-line holographic microscopy for label-free identification and tracking of biological cells},
journal     = {Military Medical Research},
volume      = {11},
number      = {1},
year        = {2024},
}

@misc{clic:21,
howpublished = {\url{https://archive.compression.cc}},
}

\end{document}